\documentclass{article}

\usepackage{amssymb}
\usepackage{graphicx}

\usepackage{url}

\usepackage{hyperref}
\usepackage{float}
\usepackage{caption}
\usepackage{subcaption}

\begin{document}
\title{Opinion dynamics: models, extensions and external effects\label{chap:modelling}}
\date{}
\maketitle
%

{\centering
  \textbf{Alina S\^irbu$^1$, Vittorio Loreto$^{2,3,4}$, Vito D.~P.~Servedio$^{5,2}$ \\and Francesca Tria$^3$}\\[3mm]  
  $^1$University of Pisa, Department of Computer Science, Largo Bruno Potecorvo 3, 50127, Pisa, Italy\\[3mm]
  $^2$Sapienza University of Rome, Phys. Dept, P.le A.~Moro 2, 00185 Rome, Italy\\[3mm]
  $^3$ISI Foundation, Via Alassio 11/C, Turin, Italy\\[3mm]
  $^4$SONY Computer Science Lab, 6, rue Amyot, 75005, Paris, France\\[3mm]
  $^5$CNR, Institute for Complex Systems, Via dei Taurini 19, 00185 Rome, Italy\\[3mm]
}
\vspace{1cm}

\abstract{
Recently, social phenomena have received a lot of attention not only from social scientists, but also from physicists, mathematicians and computer scientists, in the emerging interdisciplinary field of complex system science. 
Opinion dynamics is one of the processes studied, since opinions are the drivers of human behaviour, and play a crucial role in many global challenges that our complex world and societies are facing: global financial crises, global pandemics, growth of cities, urbanisation and migration patterns, and last but not least important, climate change and environmental sustainability and protection. 
Opinion formation is a complex process affected by the interplay of different elements, including the individual predisposition, the influence of positive and negative peer interaction (social networks playing a crucial role in this respect), the information each individual is exposed to, and many others. 
Several models inspired from those in use in physics have been developed to encompass many of these elements, and to allow for the identification of the mechanisms involved in the opinion formation process and the understanding of their role, with the practical aim of simulating opinion formation and spreading under various conditions. 
These modelling schemes range from binary simple models such as the voter model, to multi-dimensional continuous approaches. 
Here, we provide a review of recent methods, focusing on models employing both peer interaction and external information, and emphasising the role that less studied mechanisms, such as disagreement, has in driving the opinion dynamics.
Due to the important role that external information (mainly in the form of mass media broadcast) can have in enhancing awareness of social issues, a special emphasis will be devoted to study different forms it can take, investigating their effectiveness in driving the opinion formation at the population level. 
The review shows that, although a large number of approaches exist, some mechanisms such as the effect of multiple external information sources could largely benefit from further studies. 
Additionally, model validation with real data, which are starting to become available, is still largely lacking and should in our opinion be the main ambition of future investigations.   
}

\section{Introduction}

The discovery of quantitative laws in the collective properties of a large number of people, as revealed for example by birth and death rates or crime statistics, was one of the factors pushing for the development of a science of statistics in the 19th century. It let many scientists and philosophers to call for some quantitative understanding on how such precise regularities arise out of the apparently erratic behaviour of single individuals. 
Hobbes, Laplace, Comte, Stuart Mill and many others shared, to a different extent, this line of thought~\cite{ball04}. Also, Majorana in his famous tenth article~\cite{majorana42,majorana42eng} pointed out the value of statistical laws for social sciences. 
Nevertheless, it is only in the past few years that the idea of approaching society in a quantitative way has changed from a philosophical declaration of principles to a concrete research effort involving a critical mass of scientists, above all physicists.  
The availability of new large databases as well as the appearance of brand new social phenomena (mostly related to the Internet world) have been instrumental for this change.

In social phenomena the basic constituents are humans, i.e., complex individuals who interact with a limited number of peers, usually negligible compared to the total number of people in the system. 
In spite of that, human societies are characterized by stunning global regularities~\cite{buchanan07}.
We find transitions from disorder to order, like the spontaneous formation of a common language/culture
or the emergence of consensus about a specific issue and there are examples of scaling and universality as well.
These macroscopic phenomena naturally call for a statistical physics approach to social behaviour,
i.e., the attempt to understand regularities at large scale as collective effects of the interaction among single individuals.

Human behaviour is governed by many aspects, related to social context, culture, law and other factors. 
Opinions and believes are at the basis of behaviour, and can be seen as the internal state of individuals that drives a certain action. 
We hold opinions about virtually everything surrounding us, hence understanding opinion formation and evolution is key to explaining human choices. 
Opinion formation is a complex process depending on the information that we collect from peers or other external sources, among which mass media are certainly the most predominant. 
Hence, understanding how these different forces interact can give insight into how complex non-trivial collective human behaviour emerges and how  well formulated information may drive individuals toward a virtuous behaviour.

In the context of sustainability challenges, the cumulative sum of people's individual actions has an impact both on the local environment (e.g., local air or water quality, noise disturbance, local biodiversity, etc.) and at the global level (e.g., climate change, use of resources, etc.). 
It is thus important to shed light on the mechanisms through which  citizens awareness of environmental issues can be enhanced, and this is in turn tightly related to the way citizens perceive their urban environment.
In this perspective, models of opinion dynamics can be applied to investigate mechanisms driving  citizens' environmental awareness. 
Very important in this sense is the effect of the information citizens are exposed to~\cite{Gargiulo2008a}, both coming from mass media and from more personalized information, expressly tailored on individuals. 
It is then crucial to consider different modelling approaches to opinion dynamics in order to have a clear outline of the state of the art and to learn from their principles.

Traditionally studied by social science, formation of opinions, as well as other social processes, have become increasingly appealing to scientists from other fields~\cite{Castellano2009a}. 
A large amount of work is concentrated in building models of opinion dynamics, using tools borrowed from physics, mathematics and computer science. 
Typically, such models consider a finite number of connected agents each possessing opinions as variables, either discrete or continuous, and build rules to explain opinion changes, resulting from interactions either with peers or other sources. 
Although assumptions and simplifications are made in building such models, they have proven very useful in explaining many aspects of opinion formation, such as agreement, cluster formation, transitions between order (consensus) and disorder (fragmentation). 
These models can help to give insights on the dynamics of the opinion formation process and eventually to   make predictions that can  be tested and backed up by real data, in a virtuous loop where results from modelling and experiments can be integrated  and can be used to open and shed light  on new questions.

In the following, we provide a review of opinion dynamics models by classifying them
according to the presence or not of external information, that is a mechanism mimicking a sort of mass media broadcast.
In Sec.~\ref{sec:models} no information is present, while in Sec.~\ref{sec:info} the external information is taken into account as an immutable agent participating in the dynamics.
Each of the above sections are further split according to the effective form of the opinion, which can be modelled either as a one dimensional vector or as a multidimensional vector.
As a further classification, the vectors representing agent opinions can be either discrete, i.e., their components can assume a finite number of states, or continuous, i.e., with values in the domain of real numbers.
A separate section, Sec.~\ref{sec:norms}, is dedicated to a quick review of models coping with the formation and respect of social norms, a subject tightly connected to environmental issues and sustainability.

This work is by no means intended as an exhaustive review of methods, although efforts have been made to include as many contributions as possible.


\section{Existing models of opinion dynamics and extensions\label{sec:models}}

One of the first and most popular models adapted to opinion dynamics from physics~\cite{galam82,galam91} is the Ising model~\cite{baxter2007exactly}. This can be though as an extremely simplified agent based model.
In agent based models, individuals are considered as independent agents that communicate each other and update their opinions according to a limited set of fixed rules. The interaction between agents may be carried on pairwise or in groups. Agents are connected by an underlying graph defining the topology of the system and the interactions are usually between nearest neighbours. Agents are endowed with opinions, that may be represented as a variable, or a set of variables, i.e.\ represented by a vector with given components, discrete, i.e.\ that can assume a set of predefined values, or continuous.

In the Ising model, each agent has one opinion  represented as a spin, that can be up or down, determining a choice between two options. Spin couplings represent peer interactions and external information is the magnetic field. 
This may appear too reductive, thinking about the complexity of a person and of each individual position. 
Everyday life, however, indicates that people are sometimes confronted with a limited number of positions on a specific issue, which often are as few as two: right/left, Windows/Linux, buying/selling, etc. 
Further, despite its simplicity, this model is particularly attractive since it foresees a phase transition from an ordered to a disordered phase, related to the strength of the spins interaction (inverse temperature in the physics language). 

Although an interesting approach, the Ising model can be too simple to interestingly account for the complexity of each individual position and of individuals interactions. 
Hence, in the last decade, many other models have been designed (an extensive earlier review of these can be found in ~\cite{Castellano2009a}). 
The aim of these chapter is to present some of these models and a selection of their latest developments.

\subsection{One-dimensional models\label{sec:1d}}

\subsubsection{Discrete opinions\label{sec:1dd}}

\paragraph{\textbf{The voter model}\\}
\indent

\begin{figure}[t]
\centering
\includegraphics[width=0.9\textwidth]{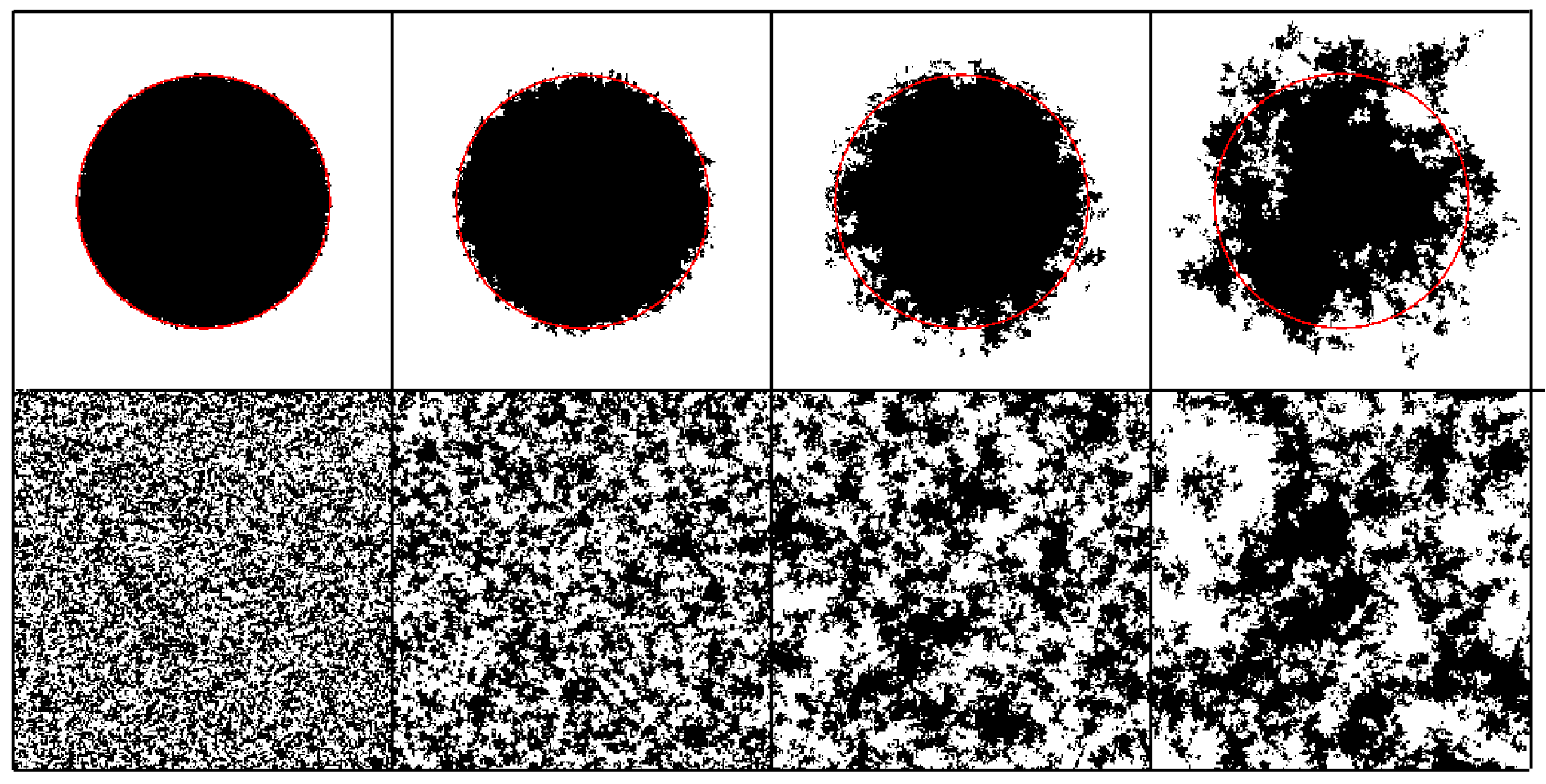}
\caption{
  Evolution of a two-dimensional voter model starting from a circle (top) or a fully disordered configuration (bottom). The white and black colours represent the positive and negative opinions respectively. From the top panel we can see how the black area remains practically constant during the dynamics and the original circular shape is destroyed. In physics, this signals a lack of surface tension.
  From~\cite{dornic01}.
\label{dornic01_1}}
\end{figure}

The voter model is one of the simplest models of opinion dynamics, originally introduced to analyse competition of species \cite{clifford73}.  
The model has then been attracting a large amount of attention in the field of opinion dynamics, and its name stems from its application to electoral competitions \cite{holley75}. 
In this model, each agent in a population of $N$ holds one of two discrete opinions, $s = \pm 1$, similar to the Ising model mentioned above.  Agents are connected by an underlying graph defining the topology of the system.
At each time step, a random agent $i$ is selected along with one of its neighbours $j$ and the agent takes the opinion of the neighbour. 
Thus, while spins in the Ising model try to align with the majority of their neighbours, voter dynamics involve one neighbour only, hence the majority does not play a direct role, but is felt indirectly through peer interaction. This difference in the updating rule is reflected in the patterns generated in two-dimensional lattices (Fig.~\ref{dornic01_1}), where domains of agents with the same opinion grow but interfaces between different domains are very rough, unlike usual coarsening systems~\cite{bray94}. 
\begin{figure}[t]
\centering
\includegraphics[width=8cm]{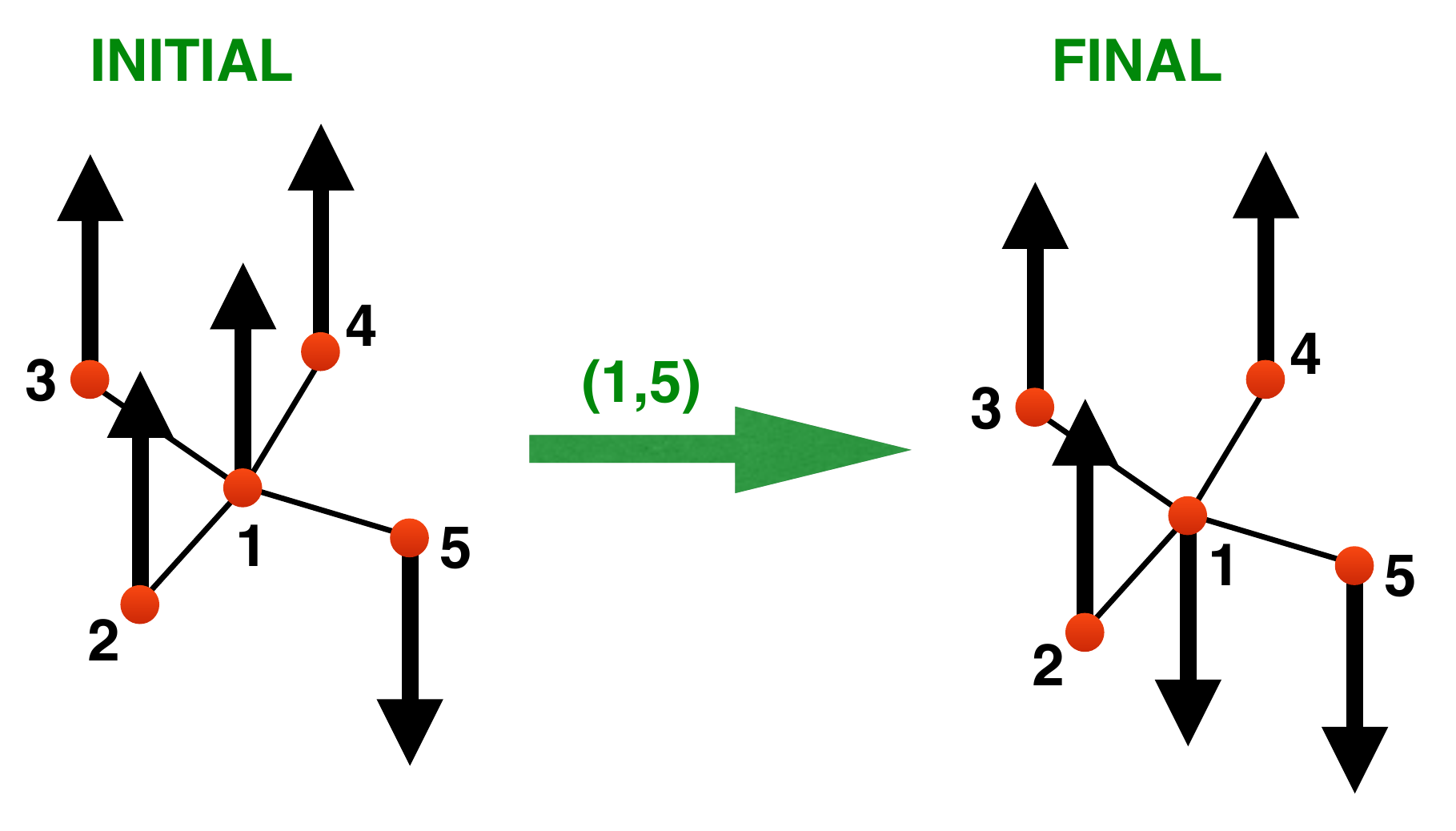}
\caption {
  Basic voter model interaction.
  Suppose that in the dynamical evolution of the model, which considers an interaction between an agent and one of its neighbours chosen at random, the agent number 1 was selected in the configuration of the left part of the figure. With probability 3/4 it will remain with a positive opinion since three of its neighbours have a positive opinion (the agents 2, 3, and 4), while with probability 1/4 it will change it since one of its neighbours has a negative opinion (the agent 5). In the example, the final state on the right refers to this latter event.
	\label{fig:voter}
}
\end{figure}
A generalized framework that encompasses different variations of voter dynamics has been introduced recently in \cite{Moretti2013a}.

The voter model dynamics has been extensively studied when  people are modelled as vertices in a  $d$-dimensional hyper-cubic lattice.
When considering a finite system, for any dimension $d$ of the lattice, the voter dynamics leads to one of the two possible consensus states: each agent with the same opinion $s=1$ or $s=-1$. The probability or reaching one or the other state depends on the initial state of the population. More interestingly, in an infinite system, a consensus state is reached only for
dimensions $d \le 2$~\cite{cox89}.
The time needed for a finite system to reach consensus  is $T_N \sim N^2$ for $d=1$, $T_N \sim N \ln N$ for $d=2$, while $T_N \sim N$ for $d>2$.  
Many generalization of the plain voter model can be considered. For instance, a level of confidence can be introduced for each opinion, determining the probability for an agent to change it.
 The confident voter model, where confidence is added to the agent state as a binary variable, converges to confident consensus in a time that grows as $\ln N$ on a complete graph, after crossing a mixed state of unsure agents \cite{Volovik2012}. On a lattice, however, consensus time grows as a power law in $N$, with some configurations crossing a long-lived striped state. 

The voter model in two dimension, with temperature, has been applied to explain opinion change in financial markets \cite{Krause2012}. The temperature (a type of noise) is associated to the nervousness of agents (fear). Through a feedback between the status of the entire agent population (market imbalance) and the temperature, nervousness becomes an evolving feature of the system. This passes through two types of metastable states, either long-lived striped configurations or shorter mean-field like states. 

A recent development involves using power-law intervals between interactions \cite{Takaguchi2011}, as opposed to the nearest neighbours or exponential interval distribution in the original model. 
The analysis is performed on different network topologies, i.e.\ ring, complete graphs and regular random graphs. 
In general, power-law intervals slow down the convergence time, with small, if no differences, seen for the complete graphs, medium for regular random graphs and large for the ring. 
The same slowing down of dynamics is shown for update probabilities inversely proportional to the time since the last change of state or interaction, which in the end lead also to power-law inter-event time distributions \cite{Fernandez-Gracia2011,Fernandez-Gracia2013}. 
However, depending on how the probabilities are defined, so called `endogenous' and `exogenous' rules, full consensus can be reached or not, respectively.

In \cite{Benczik2009}, the voter model is analysed on random networks, where links are rearranged in an adaptive manner, based on agent similarity (links with agents not sharing the opinion are dropped in favor of new connections to individuals having the same opinion). They show analytically that in finite systems consensus can be reached, while in infinite systems metastable states can persist for an infinitely long time. 
A different analysis on a \emph{directed} adaptive network has been proposed in \cite{Zschaler2011} where link directionality is shown to induce an early fragmentation in the population.

A non-linear extension of the model is introduced in \cite{Yang2011}. This allows agents to select their opinion based on their neighbours using a parameter $\alpha$ which controls the herding effect, i.e., the inclination of individuals to behave collectively as a whole. 
The probability that an agent adopts opinion $+1$ is
\begin{equation}
P(+1)=\frac{n_+^\alpha}{n_+^\alpha + n_-^\alpha},
\end{equation}
where $n_+$ ($n_-$) is the number of neighbours holding an opinion $+1$ ($-1$). For $\alpha=1$ the original voter model is retrieved, while for large $\alpha$ a model similar to the majority rule (next paragraph) is obtained. Convergence time is analysed depending on $\alpha$, and it is shown that a minimum is obtained for moderate values of $\alpha$. For extremely low values, large clusters form slowly, while for very large values, large opinion clusters take long to merge. This indicates that in order to accelerate consensus, the local majority opinion should not be strictly followed, but this should be followed in a moderate way. The optimal $\alpha$ decreases with system size. This holds for a few network types analysed, i.e.\ regular lattices, Erdos-Renyi random graphs, scale-free and small-world networks. For the complex networks, the minimum $\alpha$ is also shown to increase with the network connectivity (average degree of the nodes). 

Several other studies of non-linear dependence of an agent's opinion on the neighbours exist \cite{Schweitzer2009,Tanabe2013}. The introduction of `contrarians' has been studied in \cite{Tanabe2013}, with three types of stable states obtained: (1) coexistence of the two opinions with equal fractions, (2) adoption of one opinion by contrarians and the other by the rest of the agents or (3) a limit cycle. Zealots have been shown to prevent consensus or robust majorities even when they are in small proportion \cite{Mobilia2007}, with a Gaussian distribution of the magnetization of the system when a small equal number of zealots are added for each opinion. 

In \cite{Fotouhi2013}, the voter model with popularity bias is analysed. Here, the probability of a node to choose a particular state is not only based on the states of the neighbours, but also on their connectivity. The system is shown to reach consensus in a time $T\sim[\ln N]^2$, faster than the original voter model. When confidence is introduced, i.e.\ the probability of a state depends also on the current state of the agent, convergence to unanimity is slower. Irreversibility is also analysed, by making state $+1$ fixed, i.e.\ once agents reach this opinion they remain in that state. This is shown to converge to consensus on opinion $+1$ in logarithmic time.

Various other generalizations of the model have been proposed in the last years. An extension to three opinions has been developed in \cite{Mobilia2011}, where a third `centrist' opinion ($0$) was introduced, as standing in the middle of the two `extreme' opinions $\pm1$. Transitions from an extreme to the neutral opinion are governed by a parameter $q$ measuring the bias towards extremism, with interactions between extremists impossible (constrained voter model). The authors show that polarization is favored for $q>0$, however there is always a finite probability for consensus, while in the case of $q<0$ consensus is more probable. Addition of centrist zealots (centrists who preserve their opinion) changes the system behaviour in that a large fraction of centrist zealots generates consensus on the neutral opinion \cite{Mobilia2013}. A small zealot fraction leads to mixed populations, where centrists coexist with either both extremist types (when the two are equally persuasive) or with the most persuasive 
one.  

Strategic voting introduced in the three-state voter model \cite{Volovik2009} can reproduce patterns seen in real voting data, where two parties have similar votes and compete for the majority while the third party remains a minority over years. Stochastic effects can, however, interchange one majority party with the minority one, on a time scale growing exponentially with the size of the population, which has also been observed in real elections. 

Kinetic interaction rules for the three state model (with states $o_i\in{\pm1,0}$) have been analysed in \cite{Crokidakis2012}, where agents were influenced by two terms, a self conviction term and a peer effect term:
\begin{equation}
o_i(t+1)=C_io_i(t)+u_{ij}o_j(t)
\end{equation}
Peer interactions could be positive or negative ($u_{ij}\in{\pm1}$) while convictions could be positive, negative or missing ($C_i\in{\pm1,0}$). The probability distributions for these values determined the point of a transition between an ordered and a disordered state, with negative interactions leading to increased disorder. Real valued convictions that controlled which of the two peers will take the other's opinion and conviction were introduced in \cite{Crokidakis2013}. Noise was added to the system in the form of an instantaneous adoption of opinion $0$ by an agent in the neighbourhood of the interacting pair, with probability $p$. Stationary states were obtained for all noise levels, with one of the $\pm1$ opinions disappearing and the other coexisting with the null opinion (undecided). Very high noise led to states where most agents were undecided, while consensus states were only obtained in the noiseless case ($p=0$). An external effect was also introduced, which affected again the neighbourhood of 
the interacting pair. This was shown to decrease the number of undecided agents in the population. Also, a large strength of the external effect was shown to decrease its success. 

The voter model with arbitrary number of options was also analysed on co-evolving networks \cite{Malik2013}. An agent could either convince a neighbour of their opinion or disconnect and rewire to another agent. This rewiring was performed in a preferential manner, i.e.\ agents close in the network and flexible towards one another were selected more often. The probability distribution determining how often an agent accepted the other's opinion accounted for the `social environment', while preferential rewiring for `social clustering'. Depending on the flexibility of the social environment, two system states were observed. A flexible society evolved to a large connected component of agents sharing the same opinion (`hegemonic consensus'), with a small-world network structure. An inflexible society resulted in multiple components disconnected from the others, a so called `segregated consensus'. Within each component, agents shared the same opinion, which could be different from the other components. 
 
 \paragraph{\textbf{The majority rule (MR) model}\\}
\indent

\begin{figure}[t]
\centering
\includegraphics[width=8cm]{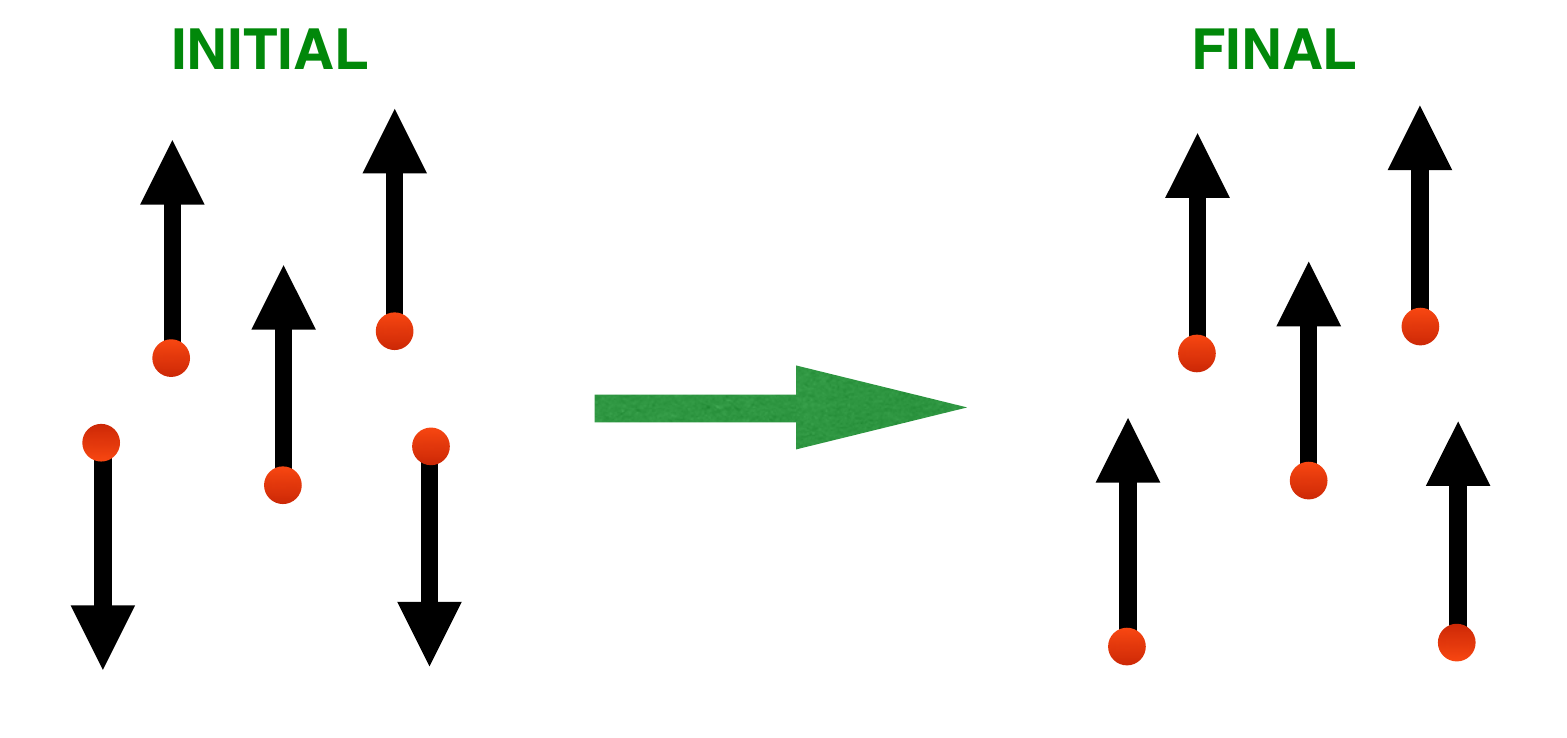}
\caption { Majority Rule model. 
	The majority opinion inside a discussion group (here of size five) is taken by all agents.
  \label{fig:majority}}
\end{figure}

The MR model was first proposed to describe public debates \cite{galam02}. Agents take discrete opinions $\pm 1$ and can interact with all other agents (complete graph). At each time step a group of $r$ agents is selected randomly and they all take the majority opinion within the group, as exemplified in Fig.~\ref{fig:majority}. The group size can be fixed or taken at each time step from a specific distribution. If $r$ is odd, then the majority opinion is always defined, however if $r$ is even there could be tied situations. To select a prevailing opinion in this case, one possibility is to introduce a bias in favor of one opinion, say $+1$. This idea is inspired by the concept of social inertia \cite{friedman84}. The MR model with opinion bias was originally applied to describe hierarchical voting in society \cite{galam86,galam90,galam99,galam00} with the discussion recently extended to three discrete choices for hierarchical voting \cite{Galam2013}.

If we define $p_+^0$ to be the initial fraction of agents with the opinion $+1$, and we allow the system to evolve, all agents will have opinion $+1$ ($-1$) if $p_+^0>p_c$ ($p_+^0<p_c$). 
If $r$ is odd, $p_c(r)=1/2$, due to the symmetry of the two opinions.
If $r$ is even, $p_c<1/2$, i.e., the favored opinion will eventually be adopted by the entire population, even if initially
shared by a minority of agents. 
To reach the consensus, the number of updates per agent scales like $\log N$~\cite{tessone04}.  
Under power-law noise, the system relaxes in a state with constant magnetization if the noise amplitude is under a threshold, while for higher amplitude the magnetization tends to zero \cite{Stauffer2008a}.

A full review of extensions and application of the MR model can be found in \cite{Galam2008a}. 
Recent extensions have been used to explain results of public debates on different issues such as global warming, evolution theory, H1N1 pandemic \cite{Galam2010}. These include two types of agents, floater and inflexible, where inflexible agents do not change their opinion. It is shown that, for the case where not enough scientific data is available, the inflexible agents are those that drive the result of the debate. Hence, a strategy for winning a debate is the acquisition of as many inflexible agents as possible. Also, the analyses indicate that a fair discourse in a public debate will most likely lead to losing, while exaggerated claims are very useful for winning. Similar results are presented in \cite{Galam2008}, where contrarians, i.e. agents who take the minority opinion of a group, are also introduced. The effect of introducing both contrarians and inflexible agents is discussed in \cite{Jacobs2008}, and 
results from the previous studies confirmed.

The same issue of public debates has been analysed with a different variation of the model \cite{Galam2010a}. Here, collective beliefs are introduced as an individual bias to select one or the other opinion, in case of a tie in voting. Here only pair interactions are analysed. The study shows that collective beliefs are very important in determining the results of the debate, and again, a winning strategy is acquiring inflexible agents, which may mean using overstated or exaggerated statements. A similar model has been also applied to explain the formation of bubble crashes in the financial market \cite{Galam2011}. Agents decide to sell or buy depending on the majority rule and the collective beliefs in case of tie. The model shows that it is the collective beliefs that determine a discrepancy between the real and the market value of an asset, which in turn generates crashes. If the collective beliefs are balanced, or ties do not appear (by using odd-sized groups), these crashes do not appear. 

Two model extensions with independent agents and collective opinions have been introduced in \cite{Wu2008}. Here, the MR model is applied with probability $1-q$, while with probability $q$ the agent either chooses one random option (extension 1) or follows the collective opinion (model 2). The authors show that, in both cases, there exists a threshold for $q$ under which complete consensus is obtained.

The majority rule model has been analytically studied on hypergraphs, with a version entitled `spatial majority rule model' \cite{Lanchier2012}. Hyperedges consisting of $n$ vertices were used to define social groups. Agents on a hyperedge simultaneously changed their opinion to that of the majority on the same hyperedge, while ties resulted in adoption of opinion $+1$. The system was shown to converge to a majority of +1 for $n$ even and to cluster for $n$ odd, even with an infinite number of hyperedges. 

A model sharing similarities to the MR model above is the non-consensus opinion (NCO) model and its extensions \cite{Li2013}. These introduce the self opinion in the majority rule, with or without a weight, and the system is shown to achieve stable states where the two competing opinions coexist, on different network types, including coupled networks.  

An application of a similar model, entitled majority vote model, to model tax evasion dynamics is presented in \cite{Lima2012,Lima2012a}. Here, $+1$ represents and honest individual, while $-1$ an individual evading tax. Individuals change their opinion with a probability which depends on the average of all of their neighbours:
\begin{equation}
P(\mbox{`flip'})=\frac{1}{2}\left[1-\sigma_i(1-2q)\, \mathrm{sign}(\sum_{n\in N(i)}{\sigma_n})\right]
\end{equation}
Here, $\sigma_i$ is the current opinion of agent $i$, $N(i)$ is the set of neighbours of $i$, while $q$ is a noise parameter. In the model an audit procedure is also introduced. When an agent chooses to evade taxes, a punishment is imposed with probability $p$, consisting in forcing the agent to be honest for a number of $k$ population updates. Different network topologies are analysed: square lattice, Barabasi-Albert and Honisch-Stauffer. Numerical results show that without punishment, tax evasion fluctuates, reaching at times very high levels. The introduction of audit, even at very low levels, is shown to reduce drastically the percentage of agents choosing to avoid tax. Similar results had been obtained previously using the Ising model \cite{Zaklan2008}. 

The majority vote model has also been analysed with heterogeneous agents \cite{Lima2013}, i.e.\ the parameter $q$ above is replaced by $q_i$, characteristic to each agent. These new parameters are drawn randomly at the beginning of the simulations from an interval $[0,q]$. Critical exponents are estimated using both analytic and numerical tools.

\paragraph{\textbf{Social impact and the Sznajd model}\\}
\indent

Interactions and opinion formation, with their complex
 underlying features, have been long analysed by social scientists, and theories devised to explain them. One example is social impact theory \cite{Latane1981}, which states that the impact
 of a group of people on an individual depends mainly on three factors: their number, their distance and their strength. A first application of this theory to build a dynamical model of opinion formation has been introduced in \cite{Lewenstein1992,Nowak1996}. This uses cellular automata to model individuals which hold one of two opinion values $\sigma_i=\pm1$. They are placed within a network, which accounts for the spatial factor, i.e. the distance $d$ between individuals. Individual strength is represented by two variables: persuasiveness (how much is an agent able to influence another) and supportiveness (how much an agent supports the opinion they hold in their neighbourhood). Social impact on individual $i$ is then computed as a weighted sum of the persuasiveness of other agents holding a different opinion and the supportiveness of agents holding the same opinion :
 \begin{equation}
 I_i=I_p\left(\sum_{j}{\frac{t(p_j)}{g(d_{ij})}}(1-\sigma_i\sigma_j)\right)-I_s\left(\sum_{j}{\frac{t(s_j)}{g(d_{ij})}}(1+\sigma_i\sigma_j)\right)
 \end{equation}
 Here $d_{ij}$ is the distance between agents $i$ and $j$ (which can be defined depending on the network type used), $g()$ is a decreasing function of $d_{ij}$ and $t()$ is a strength scaling function. Thus, the updating rule for opinion of agent $i$ is:
 \begin{equation}
 \sigma'_i=-\mathrm{sign}(\sigma_iI_i+h)
 \end{equation}
 where $h$ is a noise factor. 
The model was shown to lead to spatially localized opinion clusters, where minority clusters are facilitated by the existence of strong individuals supporting the weaker ones. This holds for a variety of social network topologies:  fully connected graph, hierarchical networks, strongly diluted networks and Euclidean space.

\begin{figure}[t]
\centering
\includegraphics[width=8cm]{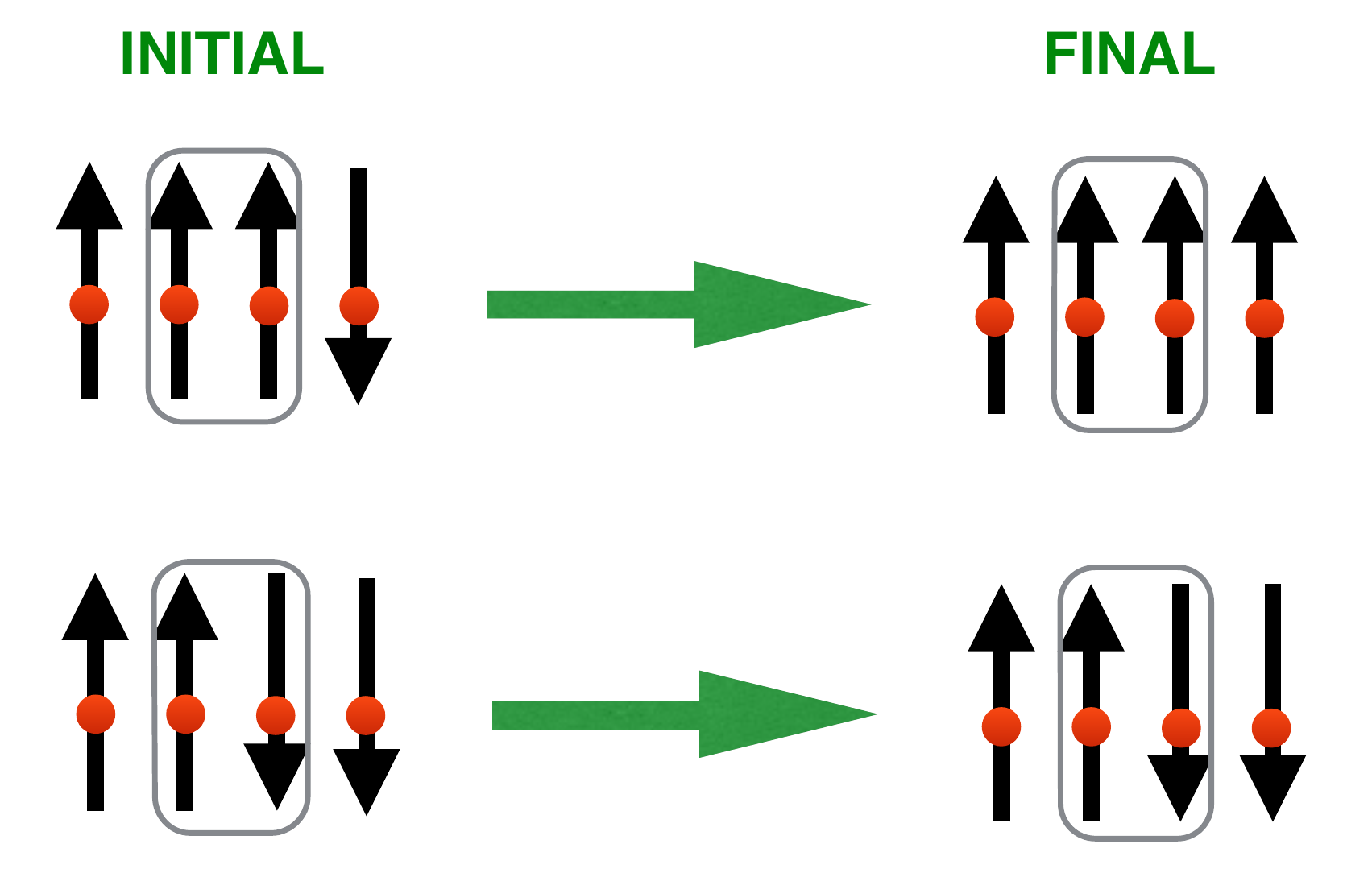}
\caption { Sznajd model. 
	A pair of neighbouring agents with the same opinion convince all their neighbours (top), while they have no influence if they disagree (bottom).
  \label{fig:sznajd}}
\end{figure}

Another recent model employing the theory of social impact is the Sznajd model \cite{Sznajd-Weron2001}. This is a variant of spin model, on a one dimensional lattice, that takes into account the fact that a group of individuals with the same opinion can influence their neighbours more than one single individual. The proximity factor is also taken into account, by considering neighbouring agents on the lattice. However, the strength of individuals, a third factor mentioned in the theory of social impact, is not present. Each agent has an opinion $\sigma_i=\pm 1$. At each time step, a pair of neighbouring agents is selected and, if their opinion coincides, all their neighbours take that opinion. Otherwise, the neighbours take contrasting opinions (Fig.~\ref{fig:sznajd}). The model has been shown to converge to one of the two agreeing stationary states, depending on the initial density of up-spins (transition at 50\% density). Versions on a two dimensional lattice have also been studied, with four neighbours (a 
plaquette) having to 
agree in order to influence their other 8 neighbours \cite{Stauffer2000}. Extensions to a third option (centrist/indifferent) have been also studied \cite{Baker2008,Malarz2009}.

A different extension is the introduction of ``social temperature'' \cite{Lama2005}. Here the original rules of the Sznajd model are applied with probability $p$, i.e.\ all neighbours take the opinion value of the plaquette, in case they agree.  With probability $1-p$ the agents take the opposite value than dictated by the original Sznajd rules. This results in disagreement by some individuals who choose to be or not to be contrarians at each update. Importantly, disagreement is not a fixed attribute of the individuals, but varies in time. It was shown that over a critical threshold for $p$, the behaviour of the original model is conserved, i.e. all individuals agree to one opinion. Under this threshold the system remains in a disordered state with magnetization (defined as ${\sum_{i=1}^N{\sigma_i}}/{N}$) close to 0.

A recent study of disagreement in the Sznajd model in one dimension is \cite{Kondrat2010}, where conformist (agreement) and anti-conformist (disagreement) reactions appear. Specifically, the model is introduced a parameter $p$ which defines the probability that, when two neighbours hold the same opinion, a third neighbour, that previously held the same opinion, will take the opposite position. If the third neighbour did not share the opinion of the initial pair, then they take that opinion, as in the original Sznajd model. It is shown that for low anti-conformity, consensus can be reached, and spontaneous shifts in the entire population between $\pm 1$ appear. On the other hand, high anti-conformity results in oscillations of the magnetization level around $0$, without reaching $\pm 1$. The same model has been applied on complete graphs \cite{Nyczka2012a}. Here, it was shown (both numerically and analytically) that the reorientations for low anti-conformism ($p$) appear now between two magnetization states $\
pm 
m$ instead of $\pm 1$. 

Agent independence (as opposed to disagreement) is studied in \cite{Sznajd-Weron2011}. Independence means that a neighbouring agent can choose not to follow an agreeing plaquette, with probability $p$. In this case, they can flip their opinion with probability $f$, described as agent flexibility. The model is analysed on one and two-dimensional lattices and on a complete graph. Independence is shown to favor coexistence of the two opinions in the society, with the majority being larger for small independence levels ($p$).

The Sznajd model with reputation, on a 2-D lattice, has been also analysed \cite{Crokidakis2011}, where each agent has a reputation value associated. The agent plaquette can influence the neighbours, with probability $p$, only if they agree and they have an average reputation larger than the neighbours. Reputations also evolve, i.e. they increase if the plaquette influences a neighbour and decrease otherwise. The model is shown to lose the phase transition for $p< p_c \sim 0.69$, when some agents preserve a non-majoritary opinion.

An analysis of the Sznajd model on an Erdos-Renyi random graph with enhanced clustering is presented in \cite{Malarz2008}, where the model is shown to not reach full consensus, unlike  the original model. The connection of a modified version of the model, which includes bounded confidence and multiple discrete opinions, with graph theory is discussed in \cite{Timpanaro2012}.

The model has been also included in a study of two competing processes, one following Sznajd and the other Voter dynamics \cite{Rybak2013}. Agents are connected by a Watts-Strogatz small-world network, and can be in two states, either S or D. At each time step, a random agent is selected. If it is in state S, it turns a random neighbour from state D to state S. If it is in state D, with probability $p$ select another random neighbour in state D, if it exists, and turn all of their neighbours into state D as well. The system is shown to switch between full consensus on S to full consensus on D depending on $p$, when the clustering coefficient is low. However as the clustering coefficient increases, the opinion S is facilitated. 

\paragraph{\textbf{The q-voter model}\\}
\indent

In \cite{Castellano2009} the non-linear q-voter model is
introduced, as a generalization of discrete opinion models. Here, N
individuals in a fully connected network, hold an opinion $\pm 1$. At
each time step, a set of $q$ neighbours are chosen and, if they agree,
they influence one neighbour chosen at random, i.e.\ this agent copies
the opinion of the group. If the group does not agree, the agent flips
its opinion with probability $\epsilon$. The voter and Sznajd models
and many of their extensions are special cases of this more recent
model. Analytic results for $q\leq3$ validate the numerical results
obtained for the special case models, with transitions from a ordered
phase (small $\epsilon$) to a disordered one (large $\epsilon$). For
$q>3$, a new type of transition between the two phases appears, which
consist of passing through an intermediate regime where the final
state depends on the initial condition. The model has been also studied on heterogeneous mean field and random regular networks \cite{Moretti2013}, where the intermediate regime is shown to disappear in the case $q>3$, behaviour qualitatively similar to that obtained on a lattice. 

In \cite{Nyczka2012} the q-voter model is analysed for non-conformity and anti-conformity with the aim to compare the two types of dynamics. Nonconformity implies that some agents, regardless of what the influencing group's opinion is, will decide to flip their opinion with probability $p$. Anti-conformity means that some agents will not follow the opinion of the group, but the opposite one, with probability $p$. The comparison shows important difference between the two types of dynamics, although they appear to be very similar. In the case of anti-conformism, the critical value $p_c$ for the order-disorder phase transition is shown to increase with $q$, while for non-conformism, this decreases with $q$.

\paragraph{\textbf{Other approaches}\\}
\indent

Binary opinions have been analysed on interdependent networks \cite{Halu2013}. Two networks were considered, each corresponding to one party running for elections. Agents were part of both networks, and chose whether to vote for one of the two parties or none based on interactions on the two different networks. A simulated annealing algorithm was used to minimize the value of a Hamiltonian that counted the conflicting connections in both networks. The method showed that the most connected network wins the elections, however a small minority of committed agents can reverse the outcome.

\subsubsection{Continuous opinions}

\paragraph{\textbf{Deffuant-Weisbuch}}
\indent

The Deffuant-Weisbuch model \cite{Deffuant2000} uses a continuous opinion space, where each individual out of a population of $N$ can take an opinion value $x_i\in[-1,1]$. Two individuals interact if their opinions are close enough, i.e. $|x_i-x_j|<d$, with $d$ a bounded confidence parameter. In this case, they get closer to one another by an amount determined by the difference between them and a convergence parameter $\mu$: 
\begin{equation}
	x_i=x_i+\mu(x_j-x_i).
	\label{dw} 
\end{equation} 
The population was shown to display convergence to one or more clusters ($c$) depending on the value of the bounded confidence parameter ($c\approx \lfloor\frac{1}{2d}\rfloor$ \cite{Carletti2006}). Parameters $\mu$ and $N$ (population size) determine the convergence speed and the width of the distribution of final opinions. A feature typical to the clusters obtained by this model is the emergence of small extreme clusters \cite{Lorenz2005}. 

The Deffuant-Weisbuch model has received a lot of attention in the
literature (see \cite{Lorenz2007a} for a previous review), with several recent analysis and extensions. For instance,
\cite{Weisbuch2002} discusses heterogeneous and adaptive confidence
thresholds on 2D lattices, while in \cite{Jager2005} the model has been extended to include disagreement in order to better describe the Social Judgment Theory \cite{Sherif1965,Griffin2012}. In \cite{Gomez-Serrano2010}, analytical
results are provided, showing that in the limit of time
$t\rightarrow\infty$, the population forms a set of clusters
too far apart to interact, at a distance larger than $d$, after which agents in individual clusters converge to the cluster's barycentre. When $N\rightarrow\infty$, the opinion
evolution is shown to be equivalent to a nonlinear Markov process, which proves the ``propagation of chaos'' for the system. This means that, as the system becomes infinite in size, an opinion evolves under the influence of opinions selected independently from the opinion process, at a rate given by the limit of the rate at which agents interact in the finite system. The initial condition and noise (`free will') were shown to have large effects on the number of clusters obtained \cite{Carro2012}. Specifically, segregated initial conditions were shown to have difficulties achieving consensus, while initial cohesion resulted in convergence to one cluster. This effect can be partially removed by noise.

The original model is based on agreement dynamics, i.e.\ if individuals
are too different, they do not interact. However, disagreement
dynamics are well known to appear in real situations \cite{Huckfeldt2004}. Hence, in
\cite{Kurmyshev2011}, partial contrarians were included, which are
agents that can disagree (i.e. change their opinion in the opposite
direction) with individuals that think differently. The society is
mixed with the two types of agents, and it is shown that dynamics
change depending on the amount of individuals that can disagree.
Depending on the value of the bounded confidence parameter, one, two
or more clusters can be observed, similar to the original
Deffuant-Weisbuch, but bifurcation patterns are different. For a large
number of contrarians, the number of clusters decreases as the
confidence increases, but clusters become more different. For a
smaller number of contrarians, on the other hand, clusters also become
closer when they are fewer. This shows that contrarians favor a more
determined fragmentation, i.e.\ not only the number of clusters, but also the distance between clusters increases. Also, the new type of agents increases the
time required to reach a final frozen state. A similar approach can be
found in \cite{Huet2008}, where the 2-D Deffuant model with
disagreement is analysed, and shown to favor extremist clusters. The model with partial contrarians presented in \cite{Kurmyshev2011} has also been extended to include opinion leaders \cite{Kurmyshev2013}. These were represented as individuals with high connectivity and fixed opinion, while the rest of the individuals were connected by a small-world network. Depending on the bounded confidence (tolerance) of the leaders, their connectivity and opinion, different patterns were shown to emerge in the system. While for a society without contrarians, tolerant leaders are more successful, in a society with contrarians this model suggests that intolerant leaders are better able to impose their views.

Noise or opinion drift has been also analysed for this model. Earlier studies introduced noise as the possibility of an agent to switch to a random opinion, with a certain probability \cite{Pineda2009}. This resulted in a transition between a disordered state, for larger noise, to formation of opinion clusters. These clusters however differed from the original model in that opinions included was not exactly the same, but a spread was visible. Also, in certain situations, spontaneous transitions between different cluster configurations were observed. Similar results were reported by \cite{Nyczka2011}, where interactions were slightly changed so that an individual can influence more neighbours at a time. The study \cite{Pineda2011} allows individuals to change their opinion in an interval centred around the previous one, instead of the entire possible range. This type of dynamics is addressed as diffusion here. The width of the diffusion interval determines how the system behaves, with a low diffusion 
favoring consensus, with a cluster which changes its centre of mass due to continuous oscillations. Large diffusion produces clusters and fluctuation patterns similar to the previous studies. 

A different extension of the model is to consider the bounded
confidence parameter as an attribute of the individuals, hence
different for each. In \cite{Lorenz2010} heterogeneous bounds of
confidence are shown to enhance the chance for consensus, since
close-minded individuals can be influenced by the more open-minded
ones (this extension has been also applied to the Hegselmann-Krause
model described in the next section). On the same lines,
\cite{Gargiulo2008} devises a method of computing the bounded
confidence threshold based on the current individual opinion, to
obtain less confidence for extremists:
\begin{equation}\label{77eq}
d_i=1-\alpha|x_i|,
\end{equation}
where $\alpha$ controls the tolerance rate. The update rule is also changed so that extremists change their opinion less:
\begin{equation}x_i=x_i+d_i(x_j-x_i)/2\end{equation}
Additionally, the social network is determined at the beginning depending on how extreme are individual opinions (extremists interact only with similar individuals, while moderated individuals can interact with a wider range, based on a segregation parameter $\beta$). Under these new conditions, it is shown that opinions converge to one large cluster when $\alpha$ is very small or $\beta$ is very large, with some small coexisting extreme clusters, while pluralism is conserved only when extremist clusters are connected enough to continue to communicate to others (large $\alpha$ and $\beta$).

Further, in \cite{Gandica2010} an analysis of the Deffuant-Weisbuch model on scale free directed social networks is presented, and the average number of final opinions is shown to be larger, when compared to undirected networks, for high bounded confidence parameter $d$ and smaller for low $d$. Also, an analysis on an adaptive network is presented in \cite{Gargiulo2010}.

The Deffuant model with bias has been analysed in \cite{Perony2012}, in a setting reaching for consensus. The bias has been introduced in the interaction rule, where changes in individuals were larger towards the bias. Also, an hierarchical interaction structure was imposed, by adding a second stage to the classical Deffuant model: once clusters are stable, each of them defines a representative, and these interact further with no bounded confidence imposed. This approach always leads to consensus, and it was shown that the effect of strong biases is reduced by using the hierarchical consensus, compared to the original dynamics. For lower bias however it was shown to be detrimental.   

Coupling of this model with a public goods game has been studied in \cite{Gargiulo2012}. Here, the `Tragedy of commons' game has been enhanced by a social interaction component. Specifically, after each round of the game, a random agent interacts with a neighbour using the update rule of the Deffuant model, if the neighbour had at least the same payoff in the last round. The opinion value so obtained represents the probability that each agent chooses one of the two possible strategies in the next round (cooperate or defect). The authors show that cooperation can be increased by adding the social component, and that the system behaviour does not change with the social network topology.

\paragraph{\textbf{The Hegselmann-Krause (HK) model}\\}
\indent

A similar model to that presented in the previous section is the HK model \cite{Hegselmann2002}. Opinions take values in a continuous interval, and bounded confidence limits the interaction of agent $i$ holding opinion $x_i$ to neighbours with opinions in $[x_i-\epsilon,x_i+\epsilon]$, where $\epsilon$ is the uncertainty. The update rule, however, differs, so that agents interact with all compatible neighbours at the same time:
\begin{equation}
  x_i(t+1)=\frac{\sum_{j:|x_i(t)-x_j(t)|<\epsilon}a_{ij}x_j(t)}{\sum_{j:|x_i(t)-x_j(t)|<\epsilon}a_{ij}},
\label{op_eq19}
\end{equation}
where $a_{ij}$ is the adjacency matrix of the graph. So, agent $i$
takes the average opinion of its compatible neighbours. Hence, this model is more suitable to model situations like formal meetings, where interaction appears in large groups, while Deffuant is better suited for pairwise interaction within large populations. 

The model has been proven to converge in polynomial time, with at least a quadratic number of steps required \cite{Bhattacharyya2013}. It is completely defined by the bounden confidence parameter $\epsilon$, facilitating its analysis. The agent population groups into clusters as the system evolves, similar to the Deffuant model, with the number of final opinion
clusters decreasing if $\epsilon$ increases. For
$\epsilon$ above some threshold $\epsilon_c$, there can only be one
cluster. The convergence to one cluster can be very slow due to appearance of isolated individuals in the middle of the opinion spectrum.
Recently, an in-depth analysis of clustering patterns,
depending on $\epsilon$ has been performed \cite{Slanina2010}, and it
was shown that there are genuine dynamical phase transitions between
$k$ and $k+1$ clusters, and that around critical values of $\epsilon$,
the dynamics slows down. The similarities and differences between the HK and the Deffuant method in the previous section have been discussed in \cite{Lorenz2005,Lorenz2007b}, by formulating the two systems as Markov Chains. This meant considering the distribution of an infinite population of agents on a finite number of opinion classes. The cluster patterns of the two models have been proven to be intrinsic to the dynamics, and the fixed points identical for the two models.

A further study on the clustering patterns
\cite{Blondel2009} proved analytically that the population in the HK
model with real opinions (not restricted to interval [0,1], but to [0,L]) and $\epsilon=1$ converges always to clusters
that are at distance larger than $1$, and provided calculation of
lower bounds of inter-cluster distance both for finite size and a
continuum of agents. The continuum version of the model considers individuals indexed by the real interval $I=[0,1]$, which have opinions in interval $[0,L]$. Hence, for $\alpha\in[0,1]$, $x_t(\alpha)\in[0,L]$ is the opinion of individual $\alpha$ at time $t$. Defining $C_x=\{(\alpha,\beta)\in I^2 / |x(\alpha)-x(\beta)|<1\}$, the update rule becomes:
\begin{equation}
x_{t+1}(\alpha)=\frac{\int_{\beta:(\alpha,\beta)\in C_{x_t}}{x_t(\beta)d\beta}}{\int_{\beta:(\alpha,\beta)\in C_{x_t}}{d\beta}}
\end{equation}
Proofs are given that during convergence, there is always a finite density of individuals between two clusters, which indicates that the model can never converge to an unstable equilibrium. This model is shown to be the limit of the original HK model, as the population size goes to infinity. Recently, the same authors have proved similar behaviour for a continuous-time symmetric version of the model, with both discrete and a continuum of agents \cite{Blondel2010}. 

Additionally, in \cite{Mirtabatabaei2011}, an
analysis of the interaction network is performed. This is dynamic in
the HK model, and evolves with the agent opinions, to reach a steady
state where the network converges to a fixed topology, as demonstrated
by \cite{Mirtabatabaei2011}. A different approach of devising
analytical results for this model is by looking at the evolution of
the distribution of opinions in the population, i.e.\ Eulerian HK model
\cite{Mirtabatabaei2012}. 

The heterogeneous version of the model, i.e.\ where the bounds $\epsilon$ are different for different agents, is analysed in \cite{Mirtabatabaei2012a} and shown to display pseudo-stable configurations, where part of the population is static. This model is compared with another version employing bounded \emph{influence} instead of bounded \emph{confidence}. Bounded \emph{influence} states that an individual $i$ is affected by an individual $j$ if $i$ is in the influence area of $j$, i.e., ($|x_i-x_i|<\epsilon_j$). This system was shown to converge faster that the original version.

\paragraph{\textbf{Other models}\\}
\indent

Apart from the above mentioned models, several other agent-based approaches have been introduced, which share similarities with the Deffuant and Hegselmann-Krause models. 
In \cite{Mas2010} a model of continuous opinions, balancing individualization versus social integration, with adaptive noise, is introduced. Depending on the noise and individualization levels, three states of the population can be obtained: consensus, individualism or preserved pluralism.

A different agent based modelling approach is \cite{Mavrodiev2012}, where the effect of social influence on the wisdom of crowds is analysed. The concept of wisdom of crowds means that the aggregated opinion of a group is closer to the truth than individual agent opinions. In this model, agents hold one continuous opinion on an issue, and interaction is modeled as the effect of the average opinion of peers. Simulation results show that the effect of social influence depends on the initial condition. Specifically, if the initial individual opinions are far from the truth, interaction has a beneficial effect, however, if they start close to the truth, social influence results in a decrease of the wisdom of crowds.

In \cite{Acemoglu2010}, a continuous opinion model with Poissonian interaction intervals and stubborn agents is introduced. These agents do not change their opinion. The model was shown to generate continuous opinion fluctuation and disagreement in the population, i.e.\ consensus is never reached.

Biased assimilation is analysed in \cite{Dandekar2013}, building upon a continuous opinion model where agents update their opinion by performing a weighted average over their neighbours. Biased assimilation means that agents tend to reinforce/extremize their opinion when shown inconclusive information about an issue. This was introduced in the model by adding a further term in the weighting procedure which depends on the current opinion of the agent. The authors show analytically that, although the model without biased assimilation does not produce polarization, even when homophily is introduced through the weights, the introduction of the new component allows for polarization to be observed. 

An approach different from those presented until now uses the Kuramoto model of coupled oscillators to describe opinion formation \cite{Hong2011}. Two types of oscillators are considered, corresponding to agents which agree or disagree to others. Disagreeing oscillators are negatively coupled to the mean field. The paper shows that, even when oscillators have the same frequency, the introduction of disagreement leads to appearance of opposite clusters, travelling waves or complete incoherence.

Models based on kinetic exchange have also been proposed for opinion dynamics \cite{Lallouache2010, Lallouache2010a}. Here an agent holds a continuous opinion $x_i \in [-1,1]$ and a conviction $\lambda_i \in [0,1]$. Upon interaction, two agents $i$ and $j$ change their opinions depending on their own and the peer's conviction:
\begin{equation}
x_i(t+1)=\lambda_i x_i(t) + \epsilon \lambda_j x_j(t)
\end{equation}
\begin{equation}
x_j(t+1)=\lambda_j x_j(t) + \epsilon' \lambda_i x_i(t)
\end{equation}
For the heterogeneous case ($\lambda_i<\lambda$), the model was shown to display breaking of symmetry for $\lambda>\lambda_c=2/3$. That is, for values under $\lambda_c$, the system maintains an average opinion of $0$, while over that, the average opinion is non null. Detailed analytic studies followed in \cite{Biswas2011,Biswas2012}.

The kinetic model has been extended to differentiate between a person's conviction ($\lambda_i$) and their ability to influence others ($\mu_i$) \cite{Sen2011}. Hence the update rule becomes
 \begin{equation}
x_i(t+1)=\lambda_i x_i(t) + \epsilon \mu_j x_j(t)
\end{equation}
The model was shown to display the same symmetry breaking with a boundary set by $\lambda=1+\frac{\mu}{2}$. Several other extensions have been recently proposed, such as introduction of positive and negative interactions (disagreement with probability $p$) \cite{Biswas2012}, or of bounded confidence \cite{Sen2012}. 

The information accumulation system (IAS) model uses the concepts of volatility ($\Delta$) and diffusivity ($\omega$) for opinion dynamics \cite{Shin2010}. The opinion $o_i$ stands in interval $[-1,1]$ and evolves as 
\begin{equation}
	o_i^{t+1}=(1-\Delta)o_i^t+\sum_{j\in N_i}\omega\, o_j^t(1-|o_i^t|)
\end{equation}
with $N_i$ the set of neighbours of $i$. The model has been employed to analyse a system of two communities connected by inter-community links, which start with two different opinions on a subject \cite{Shin2010}. The question is under what circumstances the two communities can converge to the same opinion. The maximum ratio between inter- and intra-community links for which the two communities do not show consensus is analysed, and shown to increase as the intra-community connectivity increases. This means that although general connectivity might increase, that does not mean the two communities will converge to one opinion, since the increase in inter-community links should be higher for consensus to emerge.  

\subsubsection{Hybrid models}\label{coda}

\paragraph{\textbf{The CODA model}\\}
\indent 

Continuous Opinions and Discrete Actions (CODA) are used to
model the degree of acquiring a certain discrete opinion. The original
model \cite{Martins2008a} considered two opinions +1 and -1.
Individuals are represented on a square lattice by a continuous
probability $p_i$ showing the extent of agreement to opinion +1 (with
$1-p_i$ corresponding to -1). Based on this, the
choice of the discrete opinion $\sigma_i$ is made, using a hard
threshold:
\begin{equation}
	\sigma_i=\mathrm{sign}\,(p_i-1/2).
\end{equation} 
Individuals see only the discrete opinions of others, $\sigma_i$, and change the corresponding $p_i$ based on their neighbours, using a Bayesian update rule, which favors agreement to the neighbours. This maintains the discrete public dynamics, and introduces both a means to quantify the extent of adhesion to one opinion and a memory effect (individuals do not jump directly from -1 to +1, but change their opinions continuously). The model is applied both to the Voter model of interaction, i.e. one agent interacts with one neighbour at each step, and to the Sznajd model, i.e. two neighbouring agents influence the rest of their neighbours. For both cases, the emergence of extremism even in societies of individuals that start with mild opinions at the beginning is shown. Relatively stable domains are formed within the population, which exhibit small changes after they are established. Disagreement dynamics are introduced in the model in \cite{Martins2010}, by considering part of the population as contrarians (
they 
always disagree with their peers). This has been shown to reduce agreement in the population, but at the same time to discourage extremist opinions, compared to the original model.

The model was also analysed under the assumption of migration in social networks \cite{Martins2008}, where each individual is allowed to change position, a mechanism shown to reduce the amount of extremism observed, yielding one cluster in the end. Further, in \cite{Martins2009}, a third opinion is introduced, either as `undecided' (if $p_i$ is close to $1/2$) or a real alternative (usage of three probability values, $p_i$, $q_i$, $r_i$, for the three available options). In the first case, a decrease in the amount of mild opinions is observed, but at the same time the level of extremism (the maximum absolute value of $p_i$) decreases. In the second case, there are two different analyses performed. When the third opinion is considered independent (i), the level of agreement is similar between the three options, with extremists for each. Here, for simplification, a set of assumptions about symmetry between the choices are made. When the third choice is a transition between the initial two (ii), a higher number 
of individuals adhering to the middle option is seen.

An additional analysis \cite{Martins2012} consisted of making agents also aware of the possible effect they have on the others, and discusses also the relation to other models in the literature. The concept of `trust' was introduced also \cite{Martins2013}, with agents holding an array with the probabilities that the others are trustworthy. These probabilities evolve in time, and the system was shown to reach either agreement (for higher trust) or polarization. This study also showed that agreement is reached faster than polarization. In \cite{Deng2013}, the observation range of agents is increased and so called `clustered early adopters' are introduced (i.e.\ neighbours holding the same opinion), however are shown to not have a better chance of imposing their opinion compared to randomly spread adopters.

The CODA model has been applied to the study of the adoption of theories in the scientific world \cite{Martins2012a}, where agents could support a theory or another with certain probabilities. Also, `experimentalists' were defined as agents who not interact only with peers, but can also receive information from `Nature' (an interpretation of an external information source). A fraction $\tau$ of the scientific world is made of experimentalists, and the model indicates that if $\tau$ is small it is difficult to convince the scientific world of the validity of a theory even if indicated by experiments, unless retirement is also integrated into the model (older agents are replaced with new ones with moderate opinions). Also, the small-world case was analysed and shown to increase the adoption of the correct theory.

\subsection{Multi-dimensional models}
\subsubsection{Discrete opinions}
\paragraph{\textbf{The Axelrod model}\\}
\indent

The Axelrod model for culture dynamics \cite{Axelrod1997} has been
introduced to model culture formation based on two principles, the
preference of individuals to interact with similar peers (homophily)
and the increase in similarity after an interaction appears (also termed social influence). The culture of an individual in a population of $N$ is modeled by $F$ variables
$(\sigma_1,\ldots,\sigma_F)$. Each of these can assume $q$ discrete values, $\sigma_f=0,1,\ldots,q-1$. 
The variables are called cultural {\em features}
and $q$ is the number of the possible {\em traits} per
feature. They model the different ``beliefs,
attitudes and behaviour'' of individuals. Two individuals $i$ and $j$ interact based on their position (interact with neighbours) and their corresponding overlap:
\begin{equation} o_{ij} =
\frac{1}{F}\sum_{f=1}^{F} \delta_{\sigma_f(i), \sigma_f(j)},
\end{equation} 
where $\delta_{i,j}$ is Kronecker's delta. 
The value of $o_{ij}$ is the probability to interact: one of the features with different traits
$(\sigma_f(i) \neq \sigma_f(j))$ is selected and  $\sigma_f(j)=\sigma_f(i)$ is set. Otherwise nothing happens. 
So, interaction brings individuals closer together and is more likely for similar individuals, becoming impossible for individuals sharing no trait. 
 
\begin{figure}[t]
\centering
\includegraphics[width=0.7\textwidth]{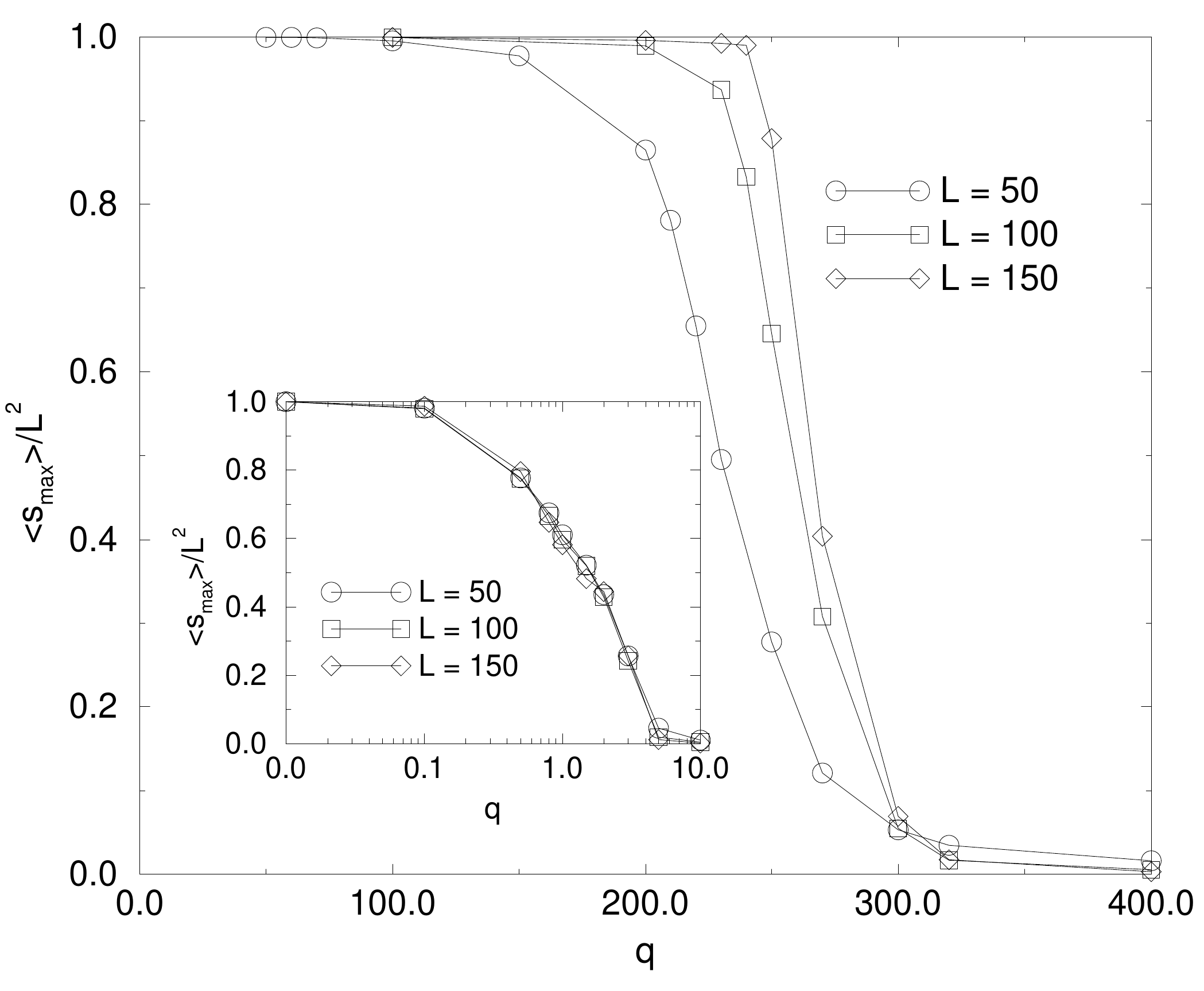}
\caption{Axelrod model. Behaviour of the order parameter $\langle
  S_{max}\rangle/L^2$ vs. $q$ for three different system sizes and
  $F=10$. In the inset the same quantity is reported for
  $F=2$. From~\cite{castellano00}.}
\label{castellano00_1}
\end{figure}

From a finite random initial population, the system evolves to one of many possible absorbing states. These can be of two types: either an ordered state, where all individuals share the same traits ($q^F$ possible states) or a frozen state where multiple cultural regions coexist. The number of possible traits $q$ in the initial population determines which of the two types of final states is obtained \cite{castellano00}. When $q$ is small, individuals share many traits so interaction and thus increasing similarity is facilitated, leading to consensus. For larger $q$, consensus is not reached because of a limited number of shared initial traits, which results in limited interaction and formation of cultural domains unable to grow. 
 On regular lattices, the phase transition between the two types of states appears at a critical value $q_c$, depending
on $F$ (Fig.~\ref{castellano00_1}).

This model has been widely analysed after its introduction, and here we present the more recent investigations. Although most studies were numerical, some analytical proofs were provided in \cite{Lanchier2010}, where it is shown that for $F=q=2$ the majority of the population forms one cluster, while a partial proof for the fact that, if $q>F$, the population remains fragmented, is provided. Also, \cite{Lanchier2013} show that for the unidimensional case, the system fixates when $F\leq cq$ where $e^{-c}=c$. Here, fixation means that the state of each individual is updated a finite number of times, until the system freezes. A similar model, designed as a generalization of the models employing homophily and influence, was introduced in \cite{Kempe2013}. Here it is shown analytically that in a system where all individuals can interact, all initial conditions lead to convergence to a stable state (invariant).  In \cite{Barbosa2009}, the dependence of the number of cultural clusters on the lattice area ($A=L^2$, 
where $L$ is the dimension of the lattice ) was analysed. They show that when $F\geq3$ and $q<q_c$, a strange non-monotonic relation between the number of clusters and $A$ exists. Specifically, the number of coexisting clusters decreases beyond a certain threshold of the area, in contrast with well known results for species-area relaxation, where the number of species increases with $A$. Outside these parameter values, however, the expected culture-area relaxation is observed. This is described by a curve that is steep at first (i.e.\ the number of clusters increases linearly with $A$) and then flattens when the maximum number of possible clusters ($q^F$) is reached. 

\begin{figure}[t]
\centering
\includegraphics[width=0.7\textwidth]{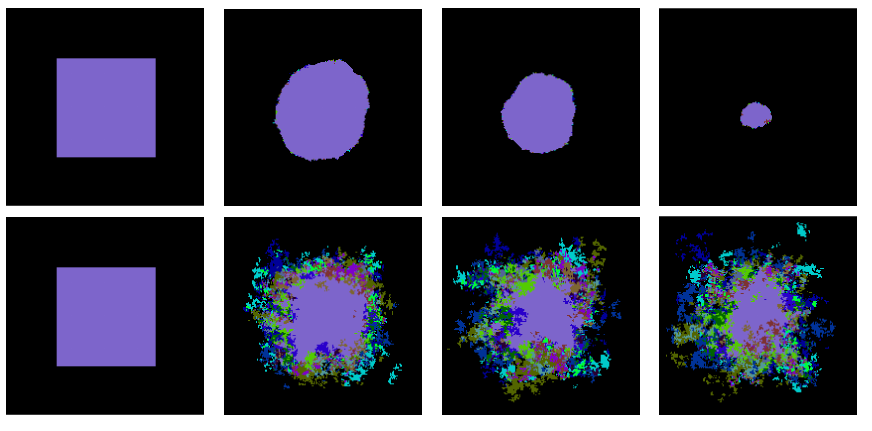}
\caption{Example of population evolution for the Axelrod model (bottom row) and the version with surface tension (top row). From~\cite{Pace2012}.}
\label{paceFig3}
\end{figure}

A recent extension \cite{Pace2012} introduced a slight modification in the updating rule, by choosing always an interacting pair of agents instead of random neighbours. This was shown to introduce surface tension in the model, resulting in metastable states for certain parameter values. Figure \ref{paceFig3} compares the dynamics of the original and surface tension Axelrod models.   

Studies of the effect of cultural drift (external noise, i.e.\ some times agents choose to change one opinion randomly) \cite{Klemm2003} showed that even a very small noise rate leads the system to agreement, while large noise favors fragmentation of cultures. Similar results were found in \cite{DeSanctis2009}, where an additional analysis of interaction noise (i.e.\ the probability to interact is modified by a small $\delta$) showed small effects on the phase transition, but a reduction of relaxation times.

Disagreement dynamics have also been introduced
\cite{Radillo-Diaz2009}, using a hard threshold for the overlap, under
which individuals disagree. Disagreement causes individuals to change
a common opinion on an issue, i.e.\ decrease their overlap. Two
different versions of this model have been developed, one where all
individuals can agree or disagree, and one where a fraction of
individuals always agrees. In both cases, disagreement dynamics are
shown to favor culture fragmentation. 

In \cite{Singh2011}, committed
individuals were introduced. These are individuals that do not change
the opinion on one of the F issues. They are introduced as a fraction $p$ of the whole population. Also, the social network evolves. The original Axelrod dynamics are changed. At each time step, an individual $i$ is selected, and one of their neighbours $j$. If $o_{ij}<\phi$, a newly defined model parameter, Axelrod dynamics are followed, otherwise, the link between node $i$ and $j$ is removed and a random node is linked to $i$. The change in consensus time due to the introduction of committed individuals is analysed. For $p=0$, consensus time grows exponentially with N, showing that rewiring impedes consensus in the population. When $p>0$, consensus time is decreased. For $p<p_c\sim0.1$, the exponential dependence is conserved, while for $p>p_c$, this becomes logarithmic in $N$. This shows that the introduction of committed individuals favors consensus in the population.

A study of the model on scale-free networks was presented in \cite{Guerra2010}. 
This analyses individuals both at ``microscopic'' - individual feature value - and ``macroscopic'' level - entire vector of features. The aim is to study how cluster composition changes when moving between the two levels. 
They show that even when many individual features are common in the population, the global culture is still fragmented.

In \cite{Banisch2010}, an application of the model to election data is
presented, using a model version with only two possible discrete
opinion values. Good similarity to election data is exhibited by the
model during the transient stage of the dynamics, i.e.\ before opinions
stabilize, when the vote distribution for each party follows the same
scaling observed in real data.

\subsubsection{Continuous opinions}
`Cultures' in the sense of the Axelrod model can be represented with continuous variables by extending continuous models like the HK to vectorial opinions (with $K$ components). Hence each position in the vector of opinions refers to a different issue. Bounded confidence dynamics lead to formation of clusters similarly to the one dimensional opinions, as shown in \cite{Fortunato2005} for two dimensions.

A different approach is presented in \cite{Lorenz2006}.
Here, the different vector elements are not independent, like in the
previous models, but they are constrained to sum to unity. In this way,
the different values could represent probabilities of choosing an
opinion out of multiple possibilities on the same issue, or could
model a resource allocation problem. The model applies bounded
confidence, by using the Euclidean distance between two individuals
($d_{ij}$). Two model versions are analysed, following
Deffuant-Weisbuch and Hegselmann-Krause dynamics. For the former,
individuals interact if $d_{ij}<\varepsilon$, when one of the peers
takes the opinion given by the average between itself and the
neighbour. Updating rules similar to the original Hegselmann-Krause
model are also defined. The model is shown to converge to one or more
clusters depending on $\varepsilon$ and $K$. When the number of
options $K$ increases, the model is shown to obtain better agreement
(large maximal component), but at the same time a larger number of
small separate clusters. Also, when $\varepsilon$ increases above a
threshold, the population converges to one opinion. This threshold
decreases with $K$. A comparison of this approach to that of considering the $K$ elements independent is provided in \cite{Lorenz2008}. In the independent case, agreement is not facilitated by an increase in $K$.

In \cite{Deffuant2012}, continuous opinions are applied to model
individuals' opinion about others and themselves, i.e.\ each individual
holds a set of N opinions. An analysis of vanity and opinion
propagation is performed, under the idea that opinions from highly
valued individuals propagate more easily. For large vanity,
individuals cluster in groups where they have a high opinion of
themselves and other group members, and low opinions of peers external
to their group. If vanity is lower, then some individuals gain high
reputation, while most of the population have a low one. Situations
with one or two agents dominating the others are exposed.

A different approach using continuous opinions and affinities between
individuals is presented in \cite{Carletti2011}. Each individual holds
a real opinion $x_i\in[0,1]$ plus a set of affinities to all other agents, i.e.\ a real vector $\alpha_i\in[0,1]^{N-1}$. These are updated simultaneously during agent interaction. The bounded confidence concept from the Deffuant model is maintained, but the definition is changed to accommodate for affinity values between individuals. Specifically, even if the opinion of two agents are not close enough, if their affinity is high, then they can still interact. Affinities, on the other hand, decrease if individuals hold diverging opinion and increase when their positions are close. The update rules are thus:
\begin{eqnarray}
x_i^{t+1}=x_i^t-\frac{1}{2}(x_i^t-x_j^t)\Gamma_1(\alpha_{ij}^t) \\
\alpha_{ij}^{t+1}=\alpha_{ij}^t+\alpha_{ij}^t(1-\alpha_{ij}^t)\Gamma_2(x_i^t-x_j^t)
\end{eqnarray} 
where $\Gamma_1(\alpha)=\frac{1}{2}[\tanh(\beta_1(\alpha-\alpha_c))+1]$ and $\Gamma_2(x)=-\tanh(\beta_2(|x|-d))$ are two activating functions that tend to step function when $\beta_1$ and $\beta_2$ are large enough. Parameters $d$ and $\alpha_c$ are the confidence thresholds, i.e.\ affinity values increase if opinions are closer than $d$, while individuals interact if their affinity is larger than $\alpha_c$. 
The model starts with random opinions and affinities, and is allowed to relax to a stable state. Affinities are then interpreted in terms of a weighted social network, with $\alpha_{ij}$ the weight of the link between agents $i$ and $j$. The authors show that the network obtained display small-world properties and weak ties.

\subsection{Modelling norms\label{sec:norms}}
Modelling norm compliance is closely related to opinion dynamics. Norms are rules enforced within society and sometimes also by law. A person can have an opinion about a norm, in the sense discussed until now, however norm compliance relates more to final behaviour, compared to opinions only. Opinions are in general indicative of behaviour, however there are cases when actions are taken in spite of contrary opinions, due to social or external pressure. Hence there are several factors to be taken into account when trying to model norm emergence, respect and violations. These start from internal predispositions and opinions, and extend to imitation of peers and, unlike pure opinion dynamics, to responses to some form of punishment.

Several agent-based approaches for building models for norm emergence and violation have appeared, many of which have a base in Game Theory. Cooperation is viewed as compliance to a norm, and defecting means norm violation. Agents hold a state that defines their strategy or probability to choose one, and change this in an attempt to maximize an utility function. This function includes different costs, punishments, rewards, etc. States are also changed based on the behaviour adopted by peers. The game theoretic literature contains many such approaches, while other hybrid agent based models have appeared recently. We give here a few recent examples of such models, to give a general idea of various approaches following these lines.     

A recent example agent-based model of norms \cite{Fent2007} shows norm evolution and coexistence in a population. Individual behaviour is represented by a continuous variable (representing the degree of adherence to a norm or another) and evolves based on in- and out-group interactions. Agents tend to be more similar to their in-group and more distant from their out-group, while being also reluctant to change behaviour. Specifically, dynamics are determined by the objective of maximizing a utility function, which includes the difference between agents and out-group, the similarity between them and in-group, and the closeness between their own behaviour at time $t$ and $t+1$ (punishment for lack of persistence). Simulation results show that when the out-group is small, the population reaches consensus to a mild behaviour, for a medium out-group several clusters form, while a large out-group results in clusters where the two extreme behaviours are acquired. This approach is very similar to opinion dynamics 
models, but adds the existence of punishment and the usage of the utility function.

In \cite{Helbing2010}, a population of $K$ individuals is divided into four types (four possible behaviours): cooperators, defectors, moralists (cooperators that punish with a cost) and immoralists (defectors that punish). Agents change their behaviour in time, based on spatial interaction with their neighbours, i.e.\ they have a larger probability $p$ to imitate their neighbour if this has a payoff $P_n$ larger that their own ($P_s$) - so called `replicator' dynamics :
\begin{equation}
p=\frac{1}{1+\exp[(P_n-P_s)/K}
\end{equation}
 The spatial effect gives an advantage to moralists, which prevail in the population, so the social norm eventually wins, with the moralists shown to profit from the presence of immoralists and defectors. 

Ignorance about norm compliance levels is discussed in \cite{Groeber2010}. The question is whether hidden norm violations can enhance or not norm compliance in general. The model includes a population of agents and an inspector agency. Agents can choose to violate or adhere to a norm, and how much effort they put in concealing a violation, while the inspector agency decides how much to invest in inspections. The chosen behaviour is derived from the publicly known number of violations plus a belief of how many undetected violations there are (a suspicion level). Various means of defining the choice and other parameters of the model are explored. The main results show that when norms are enforced by peers, ignorance reduces norm violations. However if enforcement is performed by a third party inspector, who receives awards depending on the violations discovered and punished, then ignorance actually increases the number of violations in the population. The opposite effect is explained by the competition between 
inspectors and agents. 

Recently, a study of collective behaviour \cite{Centola2013} looks at critical mass self reinforcing dynamics and how these affect stability. Willingness to participate in collective behaviour is similar to complying with a certain norm, and defines a certain agent behaviour. Here, critical mass systems are employed, where the incentive to participate increases with the number of participants (self-reinforcement). Free-riders are however still allowed, generating a so called `weak' self-reinforcement: after a certain participation level is reached, some agents might decide not to participate, since the collective behaviour is already established (incentives peak out before the collective behaviour reaches the entire population). Although in general full participation is aimed for, the authors show, using a simple threshold model, that weak self-reinforcement has the advantage of greater stability (resilience to perturbations), generating larger participation in the long run. 

A similar approach to look at norms in a public-goods setup \cite{Tessone2013a} showed the appearance of so called diversity-induced resonance in an agent based model. The model considers `conditional cooperation', a concept similar to self reinforcing dynamics, where the willingness of a user to follow a norm increases with the mass of followers. Sanctions are introduced to represent social pressure, which depend on the number of agents violating the norm and an individual `sensitivity' to this pressure. This creates diversity in the population. Additionally, the norm changes in time, by changing the effect of the social pressure. A utility function is defined using all these components plus the cost of cooperating and the gain from the public goods. Agents need to maximize their gain, using evolutionary dynamics. Two approaches are analysed, the replicator dynamics with noise and logit dynamics (compare the payoff for the two possible behaviours). The results show that indeed, norm compliance levels are 
maximized for a certain level of diversity. Similarly, an optimal range of noise levels exists, to maximise norm compliance.   

In \cite{Schweitzer2013}, cooperation is analysed in an interactive population in a prisoner dilemma setting. A heterogeneous model is employed, including aspects of the non-linear voter model (Section \ref{sec:1dd}). The strategy chosen by an agent at each time step depends not only on the previous payoff, but also on social interactions (social herding): agents would also take into account the fraction of cooperators in their neighbourhoods. The social effect was shown to facilitate the adoption of cooperation as a strategy, i.e.\ norm compliance.

A different agent based model for norm compliance has been introduced in \cite{Pietrosanto2013,Dicarlo2015}. Here agents hold a state variable determining their behaviour, i.e.\ a probability to respect or not a  norm $\sigma_i(t)$, that evolves in time. Each agent has a natural predisposition to respect a norm ($\rho_i$), and this is also the initial state. However, at each time step, agents interact in groups, randomly selected from all agents, and change their state depending on several factors. All agents tend to relax to their natural predisposition, respond to social forces (getting closer to the states of the others in the group) and can be punished (with probability $p$) in case they do not respect the norm, which makes them increase their respective $\sigma_i(t)$. In this model, no measure of payoff is used, as opposed to previous methods. Punishment is shown to increase the level of adherence to the norm, effect whose extent depends on the time scale at which the state relaxes to the natural 
predisposition. A 
`pack effect' is also introduced, where agents feel the punishment less if their group behaves the same as them. This is shown to slightly decrease the levels of norm adherence.

Norms in social (peer) production systems have been analysed in a general framework of calibration for social models \cite{Ciampaglia2013}. The emergence of such norms has been shown using a model calibrated with data from Wikipedia online communities. This links the process of norm emergence to population dynamics. Beliefs are modeled similarly to the Deffuant model, using continuous variables and bounded confidence. However, here we find two types of agents: users and pages. Users interact only with pages using Deffuant rules, i.e.\ by simulating the editing process, and in this way get an idea of the beliefs of other users. This interaction changes the state of the page and of the user. A second type of interaction is included, to model sanctions: only pages change their state, meaning that vandalism is removed with no effect on the user making the correction. The user population changes in time, with new users joining and old users retiring. Similarly, pages are created at a certain rate, and the 
selection 
of pages by users is performed based on their popularity. Indirect inference was shown to be suitable for fitting model parameters with the experimental data from the online community.

\section{Effect of external information on opinion dynamics\label{sec:info}}

The models we reviewed so far apply to situations in which consensus spreads or tries to spread among populations according to peer mutual interactions. There is no reservoir, to use a term coined in physics, with which or against which the population interacts. This limitation can be justified in few special cases, as for instance the spreading of dialects or regional behavioural habits, where the external pressure pushed on individuals comes from the interactions among the individuals themselves.
On the other hand, we are nowadays bombarded by a huge amount of external information, ``external'' meaning here that such information comes from other sources than word of mouth.
We live in a world where the mass media play a fundamental role.
In order to understand whether it is feasible to achieve whatever behavioural changes in the population in response to given stimuli, we must consider models in which there is an external source of information.
Some efforts in this direction have been made by the scientific community so far, however approaches are still limited to only a few of the models presented in the previous section. 
In the following paragraphs, we review  the state of the art of opinion dynamics modelling with external sources of information.

\subsection{One dimensional opinion}

\paragraph{\textbf{Discrete opinions}\\} 
\indent

The effect of mass media has been  studied for the Sznajd model on a square lattice \cite{Crokidakis2011a}, by introduction of an external agent (media, having value e.g. $+1$). If four neighbours agree, then all their other neighbours switch to their opinion. If they do not then the neighbours take the media opinion with probability $p$. It was shown that the final state (either all spins up or down) depends on both the initial density of up-spins and on the value of $p$. The larger $p$, the smaller the initial density of up-spins has to be to ensure full agreement to the media. For $p\gtrsim 0.18$, the population always converges to the value of the information.
In \cite{Sznajd-Weron2008}, an extension of Sznajd to three opinion states was applied to the mobile telecommunication market in Poland. The effect of media is introduced, i.e.\ an individual accepts the plaquette opinion with probability $p$, or the influence from media with probability $1-p$. Media is represented as a set of probabilities to choose one of the options. The authors found that for low advertising, small companies are taken over by larger ones, as it happens in reality.

External information with accuracy was studied for a binary opinion model in \cite{GonzalezAvella2011}. Here, the two opinion options are not equivalent and external information could take, at different time steps, one value with probability $p>0.5$ (the true or the most beneficial opinion) or the other value with probability $1-p$. At each time step a random agent was chosen to interact with this information. If the opinion of the agent was different, then it would be updated only if a fraction of the neighbours larger than a threshold $\tau$ held the same opinion as that of the external information. The system was shown to reach consensus to information only for intermediate values of $\tau$, with mixed populations with fluctuations obtained for small $\tau$, while for large $\tau$ the population froze in the initial state.   

Non equivalent binary opinions were also investigated in \cite{Laguna2013}, where agents hold either opinion $1$ or $2$, the second being the right one. Individuals update their opinions based on small group interactions where a poll decides whether to change or not. In this poll, the higher value of one opinion counts, and a weight is used for the self opinion (conviction). With probability $P$, agents can, instead of interacting with peers, interact with a so called `monitor' which forces them to adopt opinion $2$. This is one way of introducing external effects, where the persuasion of the external field is infinite. A different way is using a set of static individuals (educated group) that follow the same interacting rules as normal agents when spreading their opinion, but do not change their state. The two options are shown to increase adoption of the right option, however the educated group was less efficient than monitors.  

\paragraph{\textbf{Continuous opinions}\\} 
\indent

Effects of external information on the dynamics of the Deffuant model
have been also investigated \cite{Carletti2006}. All individuals are
exposed to an external source of information $O$, which promotes a specific opinion. Every $T$
generations, the entire population interacts with the information. These interactions follow the same rules as with
other individuals: the opinion is updated only if the bounded confidence condition is met (see Equation~(\ref{dw}) for details).  Experiments were performed with $\mu=0.5$. Dynamics
were shown to depend on the value of the information, on $T$ and on
the parameter $d$ from the original model. If the confidence is large
enough so that the information can reach all individuals, the
population converges to this. On the other hand, if confidence is
extremely small, it is shown that full agreement with the information
can be never reached. If neither of this applies, two types of
dynamics are observed:
\begin{enumerate}
\item[(i)] In the case of extreme information (close to $0$ or $1$)
  and low confidence, $T$ has to be in a fixed interval for the
  complete agreement to information to appear. Outside this interval,
  some individuals move away from the information forming an
  additional cluster. This shows that for extreme information to be
  efficient in a close-minded population, individuals need to be
  exposed to information often enough, but also need to interact to
  each other.
\item[(ii)] In case of mild information or large confidence, complete
  agreement is found only when $T$ is larger than a threshold. This shows
  that individuals that do not access the information directly (because the confidence threshold is not met) can be
  influenced only if a large number of peer interactions are allowed
  before re-exposure to information. When the population does not
  converge to the information, still, large fractions of individuals
  form a cluster around the information value (minimum value over
  $0.5$).
\end{enumerate} 

Another approach to analyzing the effects of mass media in an extension of the Deffuant model is \cite{Gargiulo2008a}, where, each generation, individuals interact with an external information $x_I$ modulated by a parameter $\epsilon$, the information strength:
\begin{equation}
x_i=x_i+\mu \epsilon d_i (x_I-x_i),
\end{equation}
where $d_i$ is defined as in Equation~(\ref{77eq}). For mild
information (low $\epsilon$), individual opinions move towards the
value of $x_I$, however for strong information, an increasing number
of antagonistic clusters emerge. This shows that aggressive media
campaigns are risky and might result in the population not acquiring
the information.

A different model similar to Deffuant's considers both disagreement
and effects of mass media (external information)
\cite{VazMartins2010}. Here, disagreement is included as an attribute
$w_{ij} \in \{-1,+1\}$ of the link between two individuals (some couples
always agree, others always disagree), and opinions take values in
interval $[0,1]$. The interaction causes a change in the opinion value
based on the type of link: \begin{equation}x_i=x_i+\mu
  w_{ij}(x_j-x_i)\end{equation} Additionally, an external information
source is considered, applied to all individuals after a specific
number of updates. The introduction of repulsive links was shown to
favor consensus with the external information.

In \cite{Hegselmann2006}, truth seekers are introduced into the Hegselmann-Krause model, i.e.\ individuals that take into account the value of the truth $T$. This can be interpreted as individuals who interact with experts, and is similar to the interaction to an external source of information. The opinion of an individual, upon interaction to a peer, changes as
\begin{equation}
x_i(t+1)=\alpha_iT+(1-\alpha_i)f_i(x(t))
\end{equation}
where $f_i(x(t))$ is the right term in Equation~(\ref{op_eq19}), while $\alpha$ represents the disposition of individuals to seek the truth (which can be seen as the strength of the information). It is important to notice that the effect of the truth is not based on bounded confidence, i.e.\ it affects individuals with $\alpha \neq 0$ regardless of their opinions. Results show that even for small $\alpha$ (0.1) for all individuals, or if at least for half of the population $\alpha \neq 0$, the population converges to the truth, provided the truth is not extreme. If the truth is extreme (close to $\pm1$), and not all agents have $\alpha \neq 0$, some individuals remain far from truth. Large values of $\alpha$ may result in more individuals with  $\alpha=0$ to stay away from truth, which means that too strong information may have a disadvantageous effect. A further analysis of the model with truth seekers is presented in \cite{Kurz2011}, where it is proven analytically that all truth seekers (individuals with 
$\alpha \neq 0$) converge to the truth, even if there are agents that do not seek the truth.

Multiple interacting mass-media sources for the Deffuant model are analysed in \cite{Quattrociocchi2013}. Here, agents are placed on a scale-free network and media sources on a complete network. Agents interact with others and the media using Deffuant dynamics. A media source interacts with the others by choosing, among its neighbours, the most successful one and getting closer to it in the Deffuant sense. Competition between media sources is also introduced by allowing disagreement between competing media sources.
The system is shown to display stable clusters of different opinions. Additionally, media competition appears to favor fragmentation in the population. 

\subsection{Multi-dimensional opinion}

\paragraph{\textbf{Discrete opinions}\\}
\indent

The effect of mass media or propaganda for the Axelrod model has been widely studied, by introducing an external agent (information source, field) that can interact with the individuals in the population. One approach is to introduce a parameter $p$ that defines the probability that, at each time step, an agent interacts with the information instead of a peer \cite{GonzalezAvella2006,GonzalezAvella2010,Peres2011,Peres2010}. In this case, it was shown that, surprisingly, a large probability to interact with the information actually increases fragmentation instead of favoring agreement. However, this could be explained by the fact that increasing the frequency of interaction to the external agent decreases the possibility of agents to interact between themselves. Hence there is an interdependence between peer and field interactions. This, coupled with the fact that, at the beginning, some individuals cannot interact with the information (low overlap), causes an isolation of these and creation of additional 
clusters.

In the above cited approaches, the external information was independent of the state of the population and never changed. Several other ways of defining external information were also analysed in \cite{GonzalezAvella2006,GonzalezAvella2007,GonzalezAvella2012}, where so called global and local endogenous fields were considered, on a two dimensional lattice. These were computed as the statistical mode of opinions either over the entire population (global) or over the neighbourhood of each agent (local), and accounted for endogenous cultural influences. These fields were also shown to facilitate segregation in the population, for large $p$, while for low $p$, cohesion and alignment to the information was observed. Quantitative differences between the types information were uncovered, with local information sources promoting uniformity in the population. Furthermore, an analysis of two separate populations, where each is influenced by a the global field of the other, has shown complex behaviour \cite{
GonzalezAvella2012}, where sometimes one population did align to the information from the others, but also could completely reject it or form a large rejecting minority.

Several methods trying to overcome the interdependency between peer and field interactions, in the quest for induced agreement, have also appeared. For example, \cite{Rodriguez2009} add a set of ``effective features'' to the individuals, i.e.\  additional values in the state vector, that are considered to be always equal to the information (mimicking in this way the way media is designed to target a social group). This causes the overlap with the information to be always non-zero, and leads to better agreement to it. Similarly, in \cite{Mazzitello2006}, a different definition of overlap to the mass media was used, again non-null for all individuals. A different approach can be seen in \cite{Rodriguez2010}. Here, the so called ``social influence'' is used, where individuals are affected by all neighbours (including the mass media, with a certain probability defined by its strength), using a procedure similar to voting. Again, this method increases the number of agents adhering to the external information with 
the increase in the media strength.
Several other model extensions have been analysed, trying to combine the effect of media with noise and social network structure \cite{Rodriguez2010,Mazzitello2006,Candia2008,Gandica2011}.

\paragraph{\textbf{Continuous opinions}\\}
\indent

In \cite{Quattrociocchi2011}, a Deffuant-like model in two dimensions, with two conflicting opinions ($x_i^1$, $x_i^2$), was studied under the effect of external influence from media and experts, on a scale-free social network. The model was applied to opinions on welfare and security. Results showed that when the media message is false, peer interaction can help the population escape the message, only if the media does not reach more than 60\% of the individuals.

A different approach to modelling continuous vectorial opinions has been introduced for a complete graph in \cite{Sirbu2013} and later for different topologies~\cite{Cucchi2015}, including disagreement and external information. Here, opinions are represented by an element in the simplex in $K-1$ dimensions, $\vec x=[p_1,p_2,\ldots,p_K]$, similar to \cite{Lorenz2006}. This can be interpreted either as an opinion on a resource allocation problem or as the probability to choose between $K$ discrete options. The model includes complex interactions, based on a similarity measure defined as the cosine overlap ($o^{ij}=   \frac{\sum_{k=1}^{K}{p_k^i p_k^j}}{\sqrt{\sum_{k=1}^{K}{(p_k^i)^2} \sum_{k=1}^{K}{(p_k^j)^2}}}$). This defined the probability that two individuals will follow agreement (opinions become more similar) or disagreement (opinions become more dissimilar) dynamics. External information (mass-media) is included as a static individual that all agents can interact with, after a peer interaction, with 
probability $p_I$. The system was shown to form one or more clusters depending 
on $p_I$, initial condition and type of information. Extreme mass-media messages and large exposure to external information proved to have a reduced success in the population, while mild messages and low exposure were more easily adopted by a large number of individuals. Full agreement with the information was obtained either for a very mild message or for a very low exposure of a non-extreme message to a compact initial configuration. The model was further developed in \cite{Sirbu2013a}, where multiple media sources were analysed and shown to lead to more realistic behaviour, i.e., stable non-polarised clusters or full agreement for external information which is not extremely mild.

\section{Final remarks}

The macroscopic properties of matter, where the forces between atoms and molecules are in principle known, can be reproduced by numerical simulations with relative success.
On the other end, the emergent collective behaviours of our societies are not easily to reproduce by numerical simulations mainly due to the fact that
human individuals are already the result of complex physiological and psychological interactions so that the social atom itself has been yet hardly understood.
The result of this uncertainty is the proliferation of modelling schemes that try to catch particular aspects of single humans and try to examine what kind of common behaviour such selected aspects might trigger or be related to.
While opinions and languages are the result of a social consensus, beliefs and awareness are somewhat more subtle since they require a sort of continuous feedback from the environment.
After an individual realizes that a small change of his/her own behaviour may lead to a better social condition, an awareness is acquired and a further phase has to follow to yield real tangible societal advantages.
The small cost in changing behaviour, e.g.\ starting trash recycling to cite one, must be sustained by a sufficiently high number of other individuals for the global advantages to be evident.
That is why modelling schemes try to exploit the conditions that lead to consensus similar to what in physics happens in phase transitions, so that an hint can be obtained on how to propel virtuous behaviours and reach the critical number of persons necessary to self-sustain the change.

With the aim to clarify the current literature on the topic of opinion dynamics, we presented an overview of recent methods for social modelling, with emphasis on less explored ingredients, e.g., disagreement between individuals and the effect of external information, which is thought to model the interaction of individuals with mass-media. 
According to how individual opinions are represented, models vary from discrete one dimensional to continuous multi-dimensional, with several types of interactions introduced. 
Although under different assumptions, many of these models have led to similar results, which agree also with some observed behaviours found in social systems.

Although lots of original models of opinion dynamics consider mostly attractive behaviour, i.e.\ two individuals sharing common interests or opinions tend to come closer to each other, the necessity to build approaches that have applicability to real settings has triggered introduction of different other types of interactions. 
Hence disagreement (contrarians), independence and zealots have been introduced in many of the models. 
These elements were shown to facilitate the coexistence of multiple opinions in all models, regardless of type. 
At the same time, the introduction more realistic interaction rules slowed down convergence to a stable state.

Noise was also analysed for most of the models discussed here, in the form of sudden shifts of opinion, which in reality can happen often. 
Low noise levels facilitate consensus both for continuous (e.g.\ Deffuant) and discrete (e.g.\ Axelrod) models. High noise, on the other hand, leads to instability of the system, either in the form of disorder or fluctuating clusters.

Different social network topologies do not have a large effect on the consensus time and qualitative structure of the final population for most models, however quantitative structure of clusters may change. When networks evolve together with opinions, consensus or cluster states can still appear. 
On the other hand, non symmetrical interaction between individuals, i.e.\ link directionality, was shown to induce fragmentation in several cases. 

The effect of external information is very important in studying real social systems, however the extent of models including this aspect is reduced. 
Previous analyses have concentrated mostly on the Axelrod and Deffuant models. For other discrete models, an external field generally causes trivial consensus to this, while a few scattered efforts have been made to analyse multidimensional continuous systems with alternative dynamics. In general, external information was shown to cause fragmentation when it is too extreme or too strong, for both discrete and continuous opinions. Mild information, analysed in the context of continuous opinions (in discrete models information is always extreme), was shown to induce cohesion in the population. This matches findings from chapter by K. Akerlof of this volume, where governmental pressure related to environmental issues was shown to have opposite
effects if not suitably expressed.

Most analyses concentrated on one static information source. However in reality information comes from multiple sources and is continuously changing. This change is many times also affected by the feedback from the population. These issues have been only slightly touched upon by the literature, and difficulties still remain in devising a framework where media and agents interact bidirectionally in a manner similar to society. 

The traces that society leaves of interactions and other effects are nowadays more and more at reach with the new communication technologies. Behaviour data can be extracted from various types of sensors, as we have seen in chapters by V. Kostakos et al., by D. Ferreira, V. Kostakos, and I. Schweizer and by Gautama et al. of this volume, but may also come from the new discipline of human computation and gaming, for which an overview was presented in this book in the chapter by V.D.P. Servedio et al. These data are enabling many analyses of social systems. 
However, when it comes to opinion formation, although conclusions from the different models appear to be realistic, application to real data is still very scarce. 
In general, outputs from discrete models have been compared to patterns seen in the data, such as strategic voter model and election output, majority vote model and debates or financial market crashes, tax evasion with majority vote and Sznajd. 
However, studies still concentrate on qualitative similarities, with a complete lack of quantitative analyses on real data.

%


\bibliographystyle{spphys}
\bibliography{./library,./oldBib}

\end{document}